\documentclass[twocolumn,nofootinbib,amsmath,prd,aps,superscriptaddress,
tightenlines,preprintnumbers,nobalancelastpage]{revtex4}

\usepackage{slashed}
\usepackage{color, verbatim}

\usepackage{latexsym}
\usepackage{amsmath}
\usepackage{graphicx}
\usepackage{amssymb}
\usepackage{graphicx}
\usepackage{bm}
\usepackage{adjustbox}
\usepackage{multirow}
\usepackage{graphicx}
\usepackage{latexsym}
\usepackage{amsmath}
\usepackage{mathrsfs}
\usepackage{enumerate}
\usepackage{adjustbox}
\usepackage{amssymb}
\usepackage{slashed}
\usepackage{graphicx}    % insert images in figures
\usepackage{caption}     % customize captions
\usepackage{subcaption}    % write subcaptions
\usepackage{wrapfig}     % wrap figures within text
\usepackage[normalem]{ulem}%provides strikeout through \sout{ }

\def\bea{\begin{eqnarray}}
\def\eea{\end{eqnarray}}

\newcommand{\bM}{{\boldsymbol{M}}}
\newcommand{\bth}{{\bar\theta}}
\newcommand{\bel}{\textcolor{blue}}

\newcommand{\bld}[1]{\boldsymbol{#1}}

\usepackage[normalem]{ulem}

\RequirePackage[colorlinks=true,urlcolor=blue,anchorcolor=blue,citecolor=blue,filecolor=blue,
               linkcolor=blue,menucolor=blue,linktocpage=true,pdfproducer=medialab,pagebackref=true]{hyperref}
\hypersetup{
  colorlinks   = true, %Colours links instead of ugly boxes
  urlcolor     = mygreen, %Colour for external hyperlinks
  linkcolor    = blue, %Colour of internal links
  citecolor   = blue %Colour of citations
}
\usepackage{orcidlink}

 \usepackage[capitalise]{cleveref}
 \crefname{section}{Sec.}{Sec.}

\usepackage{csquotes}
\definecolor{mygreen}{rgb}{0.0, 0.5, 0.0}

\usepackage[capitalise]{cleveref}

\usepackage[font=small,labelfont=bf,justification=centerlast,singlelinecheck=true]{caption}
\usepackage{mathtools}

\newcommand{\nn}{\nonumber}

\newcommand{\Tr}{\operatorname{Tr}
}

\renewcommand{\O}{{\mathcal O}}

\begin{document}

\newcount\hour \newcount\minute
\hour=\time \divide \hour by 60
\minute=\time
\count99=\hour \multiply \count99 by -60 \advance \minute by \count99
\newcommand{\mydate}{\ \today \ - \number\hour :00}

\preprint{}

\title{\Large ALP  contribution to the Strong CP problem } 

\author{V. Enguita\orcidlink{0000-0001-5977-9635}}
\email{victor.enguita@uam.es}
\affiliation{Departamento de F\'isica Te\'orica, Universidad Aut\'onoma de Madrid, \\ 
and IFT-UAM/CSIC, Madrid, Spain
}
\author{B. Gavela\orcidlink{0000-0002-2321-9190}}
\email{belen.gavela@uam.es}
\affiliation{Departamento de F\'isica Te\'orica, Universidad Aut\'onoma de Madrid, \\ 
and IFT-UAM/CSIC, Madrid, Spain
}
\author{B. Grinstein\orcidlink{0000-0003-2447-4756}}
\email{bgrinstein@ucsd.edu  }
\affiliation{Department of Physics, University of California, San Diego, USA
}
\author{P. Qu\'ilez\orcidlink{0000-0002-4327-2706}}
\email{pquilez@ucsd.edu}
\affiliation{Department of Physics, University of California, San Diego, USA
}

\begin{abstract}
We compute the one-loop contribution to the $\bar \theta$-parameter of an 
axion-like particle (ALP) with CP-odd derivative couplings. Its contribution to the 
neutron electric dipole moment is shown to be orders of magnitude
larger than that stemming from the one-loop ALP contributions to the up- and down-quark electric and chromoelectric dipole
moments.
This strongly improves existing bounds on ALP-fermion CP-odd interactions,
 and also  sets limits on previously unconstrained couplings.
 The case of a general singlet scalar is  analyzed as well.  In addition, we explore how the bounds  are modified in the presence of a Peccei-Quinn symmetry.
\end{abstract}

\maketitle
\preprint{IFT-UAM/CSIC-24-26}

%\tableofcontents

\newpage

\section{Introduction}

Axion-like particles (ALPs) are singlet pseudo-Goldstone  bosons
(pGBs) that emerge as a consequence of the spontaneous breaking of a
global  symmetry.
   They are motivated  by a variety of scenarios that aim to solve some of the  shortcomings of the Standard Model of Particle Physics (SM).
 This includes scenarios which solve the strong CP
problem~\cite{Peccei:1977hh,Peccei:1977ur,Weinberg:1977ma,Wilczek:1977pj},
theories with extra space-time dimensions~\cite{ArkaniHamed:1998pf,Dienes:1999gw,Chang:1999si,DiLella:2000dn},
  some dynamical flavour theories~\cite{Davidson:1981zd,Wilczek:1982rv,Ema:2016ops,Calibbi:2016hwq}, dark matter models~\cite{Abbott:1982af, Dine:1982ah,Preskill:1982cy}, 
  scenarios which explore a dynamical origin for Majorana neutrino
  masses~\cite{Gelmini:1980re} and string theory
  models~\cite{Cicoli:2013ana}, among others. Consequently,
 the study of ALPs is currently under intense investigation, marked by a surge in experimental proposals and robust theoretical endeavors.  The most prominent among those pGB candidates is the QCD axion~\cite{,Peccei:1977hh,Peccei:1977ur,Weinberg:1977ma,Wilczek:1977pj}, which aims to explain the absence of CP violation in the strong sector by dynamically relaxing the $\bar \theta$-parameter to the CP conserving point.

Flavour preserving, low energy CP-odd  observables are predicted by the SM to be very small, since they are suppressed by a combination of small quark masses and  small CKM mixing angles, and first arise  at multi-loop level in the perturbative expansion.
Hence they are powerful probes of new sources of CP violation. Prime among these observables is the electric dipole moment (EDM) of the neutron (nEDM),  whose dominant contribution in the SM arises from penguin diagrams involving spectator quarks~\cite{Gavela:1981sk,Khriplovich:1981ca} and is estimated to be several orders of magnitude smaller than the current experimental bound~\cite{Abel:2020pzs,Pendlebury:2015lrz}.

In this work, we focus on the case of a generic ALP (i.e. with shift-symmetric couplings to SM fermions). 
Inspired by the QCD axion, most of the effort in ALP phenomenology has
been devoted to study the CP-conserving couplings of ALPs to  SM
particles. 
However, CP-violating ALP signatures are gaining increased interest 
 \cite{DiLuzio:2020oah,OHare:2020wah,Dekens:2022gha,DiLuzio:2023edk,Plakkot:2023pui,Bonnefoy:2022rik,Grojean:2023tsd,Bauer:2021mvw,DiLuzio:2023lmd}.

 Indeed,  derivative ALP-fermion interactions can have  CP-odd couplings, and it turns out that they  may contribute to  quark 
 electric dipole moments (qEDMs) at one loop. 
  These  and the resulting contribution to the nEDM have been  computed elsewhere~\cite{DiLuzio:2020oah}.
  In this work, we compute instead the one-loop contribution to $\bar \theta$ of an ALP with CP-odd derivative couplings to the SM quarks. We will
  show that its contribution  to the nEDM is parametrically distinct than that of quark electric dipole moments and can be numerically much more important.  For completeness, the contribution to $\bar \theta$   of a generic scalar is also computed and compared.
\begin{figure}
    \centering
    \begin{subfigure}{0.49\textwidth}
        \includegraphics[width = \textwidth]{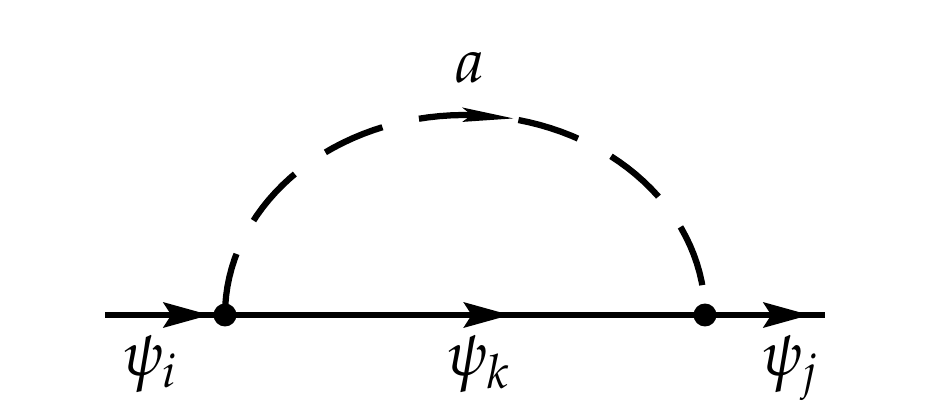}
    \end{subfigure}
    \caption{One loop corrections to the fermion masses.}
    \label{fig:1-loop-diags}
\end{figure}
 \section{Set-up: Derivative fermionic ALP Lagrangian}
 Let us consider an ALP $a$, defined as a spin-0 field,\footnote{ALPs
   with an exact shift symmetry are not necessarily pseudoscalars
   since the allowed CP-odd  derivative couplings can be as large as
   the CP-conserving ones, preventing us from assigning a definite
   transformation property of the ALP under CP.} singlet
 of the SM, and described by a Lagrangian invariant under the shift
 symmetry $a \rightarrow a + \textrm{constant}$,\footnote{ Anomalous couplings that break the shift invariance are
    often included in the ALP Lagrangian. In this work we
   ignore them.  }
  except for a small mass term $m_a\ll f_a$ where $f_a$ is the ALP physics scale. We focus here on
 effective ALP-fermion interactions up to $\mathcal {O} (1/f_a)$
 terms.  At scales $\mu\lesssim f_a$, these purely derivative interactions are
 encoded in the following effective Lagrangian:
\begin{multline}
\mathcal{L}_{a} \supset \frac{1}{2} \partial_\mu a \partial^\mu a-\frac{1}{2} m_a^2 a^2+
 \big(\bar{u}_L \boldsymbol{M}_u u_R +\bar{d}_L \boldsymbol{M}_d d_R + \text{h.c.}\big) \\
 +\bar\theta_0 \frac{\alpha_s}{8\pi} G_{\mu\nu}\widetilde G^{\mu\nu}+\frac{\partial_\mu a}{f_a} \big(
 \bar{u}_L \gamma^\mu \boldsymbol{C}_Q u_L + 
 \bar{u}_R \gamma^\mu\boldsymbol{C}_{u_R} u_R \\+
 \bar{d}_L \gamma^\mu \boldsymbol{C}_Q d_L + 
\bar{d}_R \gamma^\mu \boldsymbol{C}_{d_R}d_R\big),\label{Eq:Initial Lagrangian} 
\end{multline}
where $m_a$ is the ALP mass, $\boldsymbol{C}_{Q,u_R,d_R}$ are
hermitian $3\times3$  matrices in flavour space and $\boldsymbol{M}_{u,d}$ are the
up-type and down-type quark  $3\times3$ mass matrices. Note that the ALP couplings
to $u_L$ and to $d_L$ are identical as mandated by the SM electroweak
$SU(2)\times U(1)$ gauge invariance.  {\it Because of the hermiticity of the ALP-fermion coefficient matrices, only the flavour off-diagonal ALP-fermion couplings can lead to observable effects of CP-violation}.
 The SM fermion mass terms are
also shown in Eq.~(\ref{Eq:Initial Lagrangian}) and taken to be real and diagonal, so that the overall phase of the argument of the determinant of the mass matrices is already included in the $\bar\theta$-term, where $\bar\theta_0\equiv\bar\theta(\mu=\mu_0)\big|_{\rm tree}$  and $\mu_0\equiv f_a$.

Our objective in this work is to compute the leading contribution of the fermionic ALP couplings  to $\bar \theta$ (and its subsequent
impact on the nEDM), which arises at one loop, as we will see next; see \cref{fig:1-loop-diags}. \footnote{It has been recently claimed~\cite{Valenti:2022uii,Banno:2023yrd} that there are small additional contributions  that however are not included in the standard approach to the calculation of $\bar \theta$. The inclusion of this effect would not significantly modify the results of this work. }
 We assume that the ALP mass is larger than the QCD scale, $m_a\gtrsim 1 \, \text{GeV}$. Let us consider an
effective field theory (EFT) below the scale $\mu_2=\text{min}(m_t,m_a)$. The parameters in this EFT are computed by
matching at the scale $\mu=\mu_2$, and receive contributions from
integrating out both the ALP and the top quark.  Let us focus first on the contribution to the phase of the up quark mass. 
The details of the computation of the one-loop diagram  in \cref{fig:1-loop-diags} are given in the next section, but  let us
anticipate that the contribution to $\bar\theta$ will turn out to be dominated by the top loop, provided all entries of
$\boldsymbol{C}_{Q,u_R,d_R}$ are of the same order, and reads
\begin{align}
 \label{Eq:thetaFromALPdominantTOP}
\Delta\bar \theta_{\rm ALP}\simeq  \frac{m_t\, \text{max}(m_a^2,m_t^2)}{16\pi^2 f_a^2m_u} \text{Im}(\boldsymbol{C}_Q^{13}\boldsymbol{C}_{u_R}^{*13}) \,.
\end{align}
The dependence of this formula on the top mass is highly relevant, e.g. for $m_a<m_t$ it is cubic, a behaviour that will be transmitted to  the ensuing contribution to the nEDM. This  $\bar\theta$-induced contribution to the nEDM will be shown to be several orders of magnitude larger than that obtained in Ref.~\cite{DiLuzio:2020oah} through the contribution to the quark EDMs, as the latter only depends on $m_t$ linearly.

The mass dependence in \cref{Eq:thetaFromALPdominantTOP} can be
readily understood  using the chirality-flip basis (described in App.~\ref{App:ChiralityFlip}) instead of the derivative basis in \cref{Eq:Initial Lagrangian}. In the  chirality-flip basis, there are two relevant Feynman diagrams to be considered, shown in \cref{fig:1-loop-diags-CF}.  In order for the upper diagram in \cref{fig:1-loop-diags-CF} with an internal top quark to contribute to the $\bar u_L u_R$ term, a mass insertion $\propto m_t$ inside the loop provides the required chirality flip. Each vertex contributes a factor of $m_t/f_a$. Since the loop depends on the renormalization scale only logarithmically   the diagram gives a contribution that scales with $m_t$ and $f_a$ as $m_t^3/f_a^2$. The loop in the second diagram in \cref{fig:1-loop-diags-CF} is quadratic in $m_a$ and the vertex contributes a factor of $m_t/f_a^2$. So the contributions of the first and second diagram scale as $m_t^3/f_a^2$ and $m_tm_a^2/f_a^2$, respectively, and both contain a factor of $\boldsymbol{C}_Q^{13}\boldsymbol{C}_{u_R}^{*13}$.  Finally, since we are computing the contribution to the phase of the up quark mass term, the result is inversely proportional to the absolute value of the quark mass itself, explaining the $1/m_u$ factor in \cref{Eq:thetaFromALPdominantTOP} (see footnote 2).

The result can also be understood in the derivative basis  in
\cref{Eq:Initial Lagrangian}, in which only the diagram in
\cref{fig:1-loop-diags} contributes, but rather than obtaining the
factors of $m_t^2$ and $m_a^2$ from the vertices, they arise from the
quadratically divergent 
loop integral. 
We have verified that both bases give
the same results for the explicit computations presented
in \cref{computation}.

It is easy to see as well that the contributions to other 
CP-violation
observables do not scale as $m_t\text{max}(m_t^2,m_a^2)$. Take
for example the contribution to the quark EDM, $d_q$, and use again the
chirality-flip basis of the Lagrangian.  At one-loop it arises exclusively from
the diagram in \cref{fig:1-loop-diagsupEDM}; there is no
contribution from the analogue of that for the up quark mass from the bottom diagram in \cref{fig:1-loop-diags-CF} because the photon cannot couple to
the neutral axion. The chirality flip and the vertices give a factor
of $m_t^3/f_a^2$, but now dimensional analysis dictates that the loop
integral gives an additional factor of $1/m_t^2$, so that $d_q\sim m_t/f_a^2$.

The Lagrangian in \cref{Eq:Initial Lagrangian} includes all operators that contribute at order $\O(1/f_a^2)$ to  $\Delta \bth_{\rm ALP}$ in the derivative basis. There exist a single shift-symmetric operator with mass dimension 5~\cite{Georgi:1986df,Brivio:2017ije,Gavela:2019wzg}  and another one with dimension 6~\cite{Weinberg:2013kea,Bauer:2022rwf}, both involving the Higgs. They are not displayed in \cref{Eq:Initial Lagrangian} since they are CP-even and thus do not contribute to $\Delta \bth_{\rm ALP}$. Beyond these, there are CP-odd operators with mass dimension 7 and thus $\mathcal{O}(1/f_a^3)$. 

It is also worth noting that, since we are assuming $m_a\gtrsim 1\, \text{GeV}$, including the operator $aG\tilde G$ in the Lagrangian would not introduce a  dynamical mechanism relaxing $\bth$ to zero. That is, it would not amount to promoting the ALP to a QCD axion with a true PQ symmetry (that would require $m_a\to 0$).

\section{\texorpdfstring{Computation of the contribution to $\bar \theta$}{} }
\label{computation}
We perform in this section the one-loop matching of the Lagrangian in \cref{Eq:Initial Lagrangian} to the following QCD Lagrangian:  
\begin{multline}
\mathcal{L}_{\rm QCD,CPV}= \bar\theta \frac{\alpha_s}{8\pi} G_{\mu\nu}\widetilde G^{\mu\nu}  + \sum_{q=u,d,s}\bar{q} \, m_q q\;-\\
-\frac{i}{2} \sum_{q=u,d,s}\left[d_q F_{\mu\nu}\;\bar{q} \sigma^{\mu\nu}\gamma_5 q 
+g_s\,\tilde{d}_q\,G_{\mu\nu} \;\bar{q}\sigma^{\mu\nu}\gamma_5 q\;\right] +\cdots
\label{Eq:QCD Low energy Lag} 
\end{multline}
where $q=u,d,s$ denote the three lighter quark
fields\break  (up-, down- and strange-quarks), while $d_{u, d, s}$ and
$\tilde{d}_{u, d, s}$ stand for  their respective qEDMs and chromo-EDMs (cEDMs), and $F_{\mu\nu}$ and  $G_{\mu\nu}$ denote
the QED and QCD field strength tensors, respectively.
Other dimension five operators (such as the Weinberg operator) are left
implicit in the ellipsis.

 The calculation of $\bar\theta$ proceeds in standard effective field theoretical  fashion: starting from the EFT
  defined in \cref{Eq:Initial Lagrangian}, we run $\bar\theta$ from $\mu_0$ to $\mu_1\approx\text{max}(m_t,m_a)$, and then match $\theta$ to 
  $\theta'$ defined in EFT${}'$, a new effective theory  where the heaviest of  the top quark  and the ALP  has been integrated out.  Then run
   $\bar\theta'$ to $\mu_2$ where the lighter of the top quark  and the ALP is integrated out and a new $\theta''$ for a new EFT${}''$
  is computed. This effective theory is  QCD+QED with 5 flavours of quarks. To perform the calculation we find  the renormalization group equation
  (RGE) in the EFT, use it to determine the  functional form of $\bar\theta(\mu)$, giving $\bar\theta(\mu_1)$ in terms of the initial (prescribed)  value 
$\bar\theta(\mu_0)$, and then compute a threshold correction $\delta\bar\theta(\mu_1)$ that fixes the parameter in the EFT${}'$: 
$\bar\theta'(\mu_1)=\bar\theta(\mu_1)+ \delta\bar\theta(\mu_1)$. The process is then repeated in  going from
EFT${}'$ to EFT${}''$. Finally the RGE is used in the EFT${}''$ to  compute $\bar\theta''(\mu_{IR})$ where the choice $\mu_{IR}\approx1$~GeV is
common as it is  appropriate for computing physical effects such as the nEDM. Henceforth we  drop the double prime in
$\bar\theta''(\mu_{IR})$. If the ALP is lighter than the $b$ quark the procedure above is accordingly modified.

At leading order the ALP contribution to both the running and matching of $\bar \theta$  can be obtained using
  $\bar \theta=\theta+\arg\det(\boldsymbol{M}_u ^{\rm 1\,loop}
  \boldsymbol{M}_{d}^{\rm 1\, loop})$ by evaluating the ALP-exchange contribution to the one-loop quark mass matrices,
\begin{align}
\boldsymbol{M}_{u,d}^{\rm 1\,loop}=\boldsymbol{M}_{u,d}+\Delta \boldsymbol{M}_{u,d}\,,
\label{Eq:1loopMassmatrices}
\end{align}
where $ \boldsymbol{M}_{u,d}$ is the tree level mass and $\Delta
\boldsymbol{M}_{u,d}$ is the correction at one loop, thus
\begin{align}
\Delta \bth_{\rm ALP}(\mu)
&=\sum_{q=u,d}\arg \left[\operatorname{det} \left(\bM_{q}\left(1+\bM_{q}^{-1} \Delta\bM_{q}\right)\right)\right] \nn\\
&=\sum_{q=u,d}\operatorname{Im}\left(\operatorname{Tr} \log \left(1+\bM_{q}^{-1} \Delta\bM_{q}\right)\right) \nn
\\
&\simeq\sum_{q=u,d} \operatorname{Im}
\operatorname{Tr}\left(\bM_{q}^{-1} \Delta\bM_{q}\right)\,.
\label{eq:thetaLP}
\end{align}
Here, as stated earlier, $\bM_q$ in \cref{Eq:Initial Lagrangian} is taken to be real, and in
the last step we assumed that the loop correction to the mass is
small,  $\bM_{u,d}\gg \Delta \bM_{u,d}$.\footnote{This approximation breaks down for $m_u\to 0$, which is why the apparent divergence in
  \cref{Eq:thetaFromALPdominantTOP} in that limit is an
  artifact.}

We have computed the one-loop diagram in \cref{fig:1-loop-diags}.\footnote{The renormalization of the kinetic terms does not contribute to $\bar\theta$
  \cite{Ellis:1978hq}.} The final result using dimensional regularization and the $\overline{\text{MS}}$ scheme reads
\begin{align}
\Delta \bth_{\rm ALP}(\mu)\simeq
 \frac{1}{f_a^2}\sum_{q=u,d} \operatorname{Im} \text{Tr} \left[\bM_{q}^{-1} \mathbf{C}_{Q} \boldsymbol{L} \mathbf{C}_{{q_R}}\right] \,,
 \label{Eq:deltathetaALPresult1Finite}
\end{align}
where $\boldsymbol{L}\equiv\text{diag}(L_1,L_2,L_3)\,$ and
\begin{align}
L_k=\frac{m_{q_k}}{16
     \pi^2}\left[\left(m_a^2+m_{q_k}^2\right)\right.&\left(1+\log
     \frac{\mu^2}{m_a^2}\right)\nonumber\\
  &\left.+\frac{m_{q_k}^4}{m_{q_k}^2-m_a^2} \log \frac{m_a^2}{m_{q_k}^2}\right]\,,
 \label{Eq:deltathetaALPresult2Finite}
\end{align}
which explicitly depends on the renormalization scale $\mu$ and also
includes the finite contributions. In this expression $q_k$ stands for the different type of quarks, i.e. $\{u_1, u_2,u_3\}\equiv \{u, c,t\}$ and  $\{d_1, d_2,d_3\}\equiv \{d,s,b\}$. The running of $\bth(\mu)$  is determined by the renormalization group equation,
\begin{multline}
  \frac{d\bar\theta}{d\mu}=\sum_{q=u,d} \text{Im}\frac{d}{d\mu}\ln\det\boldsymbol{\mathcal{M}}_{q}
                          =\sum_{q=u,d}  \text{Im}\frac{d}{d\mu}\Tr\ln\boldsymbol{\mathcal{M}}_{q}\\
=   \sum_{q=u,d} \text{Im}\Tr \left(\boldsymbol{\mathcal{M}}_{q}^{-1} \frac{d}{d\mu}\boldsymbol{\mathcal{M}}_{q}\right)
\end{multline}
where, to leading order,
$\boldsymbol{\mathcal{M}}_{u,d}\equiv \boldsymbol{M}_{u,d}^{\rm
  1 \, loop}$ as given  in \cref{Eq:1loopMassmatrices}. 
Taking into account that the tree-level contribution $\bM_{u,d}$ is
$\mu$-independent, it follows that 
\begin{equation}
  \label{eq:thetaLP}
\mu \frac{d\bar\theta}{d\mu}
\simeq\sum_{q=u,d} \operatorname{Im}
\operatorname{Tr}\left(\bM_{q}^{-1} \mu\frac{d}{d\mu}\Delta\bM_{q}\right)\,, 
\end{equation}
 which, given Eq.~(\ref{Eq:deltathetaALPresult1Finite}), leads to 
\begin{equation}
\mu \frac{d\bar\theta}{d\mu}\simeq
 \frac{1}{f_a^2}\sum_{q=u,d} \operatorname{Im} \text{Tr} \left[\bM_{q}^{-1} \mathbf{C}_{{Q}} \boldsymbol{\mathcal{L}} \mathbf{C}_{{q_R}}\right] 
 \label{Eq:deltathetaALPresult1}
\end{equation}
where $\boldsymbol{\mathcal{L}}\equiv\text{diag}(\mathcal{L}_1, \mathcal{L}_2, \mathcal{L}_3)\,,$ and
\begin{align}
\mathcal{L}_k=\frac{m_{q_k}}{8 \pi^2}\left(m_a^2+m_{q_k}^2\right)\,.
 \label{Eq:deltathetaALPresult2}
\end{align}

  Neglecting threshold corrections, that is, dropping the matching  contribution  $\delta\bth$ at $\mu_1$ and $\mu_2$, the result of the calculation is
{\fontsize{9.4pt}{11pt}\selectfont
\begin{align}
\bth(\mu_{\rm IR})
&\simeq \bth_0+ \label{final-result-theta-BB}
\\
&\hspace{-0.75cm}\sum_{ {u_i}=\{u,c,t\}} \!\!\!\!\frac{m_{u_k}\,
                       (m_a^2+\widehat m_{u_k}^2)}{16\pi^2 f_a^2m_{u_i}} \text{Im}(\boldsymbol{C}_Q^{ik}\boldsymbol{C}_{u_R}^{*ik}) \log \frac{f_a^2}{\texttt{max}(m_a^2,m_{u_k}^2)}\nn\\ 
&\hspace{-0.75cm}+  \sum_{ {d_i}=\{d,s,b\}} \!\!\!\! \frac{m_{d_k}\, (m_a^2+\widehat m_{d_k}^2)}{16\pi^2 f_a^2m_{d_i}} \text{Im}(\boldsymbol{C}_Q^{ik}\boldsymbol{C}_{d_R}^{*ik}) \log \frac{f_a^2}{\texttt{max}(m_a^2,m_{d_k}^2)}\,.\nn
\end{align}}
In this expression we have included the leading log resummation  capturing the effect of QCD, as this is needed to account properly
  for the $\mu$-dependence of the quark masses as explained in \cref{app-running}. This is encoded in the
  modified masses $\widehat m_{u_k}$ and $\widehat m_{d_k}$, as given  in  \cref{hat-mass-given}, and implicitly in computing the ratios of masses as $m_{u_k}/m_{u_i}=\overline m_{u_k}(\mu)/\overline m_{u_i}(\mu)$ or $m_{d_k}/m_{d_i}=\overline m_{d_k}(\mu)/\overline m_{d_i}(\mu)$, that is, a ratio of running masses at a common renormalization scale $\mu$, cf~\cref{eq:runmass}. 
 As discussed after \cref{Eq:thetaFromALPdominantTOP}, under the assumption that the combinations $\text{Im}(\boldsymbol{C}_Q^{ik}\boldsymbol{C}_{{u,d}_R}^{*ik})$ are comparable, the dominant contribution to $\bth$ arises from the top loop affecting the phase of the up quark.

Finally, note that Eqs.~(\ref{Eq:deltathetaALPresult1Finite}) and (\ref{Eq:deltathetaALPresult2Finite}) produce in addition finite contributions to $\bth$ which add to the putative initial value $\bth (\mu_0=f_a)$. Assuming no cancellations, all contributions need to 
separately comply with the experimental limit $\bth \ll 10^{-10}$.\footnote{ This implies that the initial value of
  $\bth(\mu_0)$ also needs to be small. The bounds on the ALP couplings slightly differ depending on the mechanism that may be
  responsible for the solution to the strong CP problem at a high scale. If the complete physical $\bar \theta$-parameter is set
    dynamically to be small in the UV, $\bth (\mu_0)\ll 10^{-10}$, a non-zero $\bth$-parameter is generated solely through the running
    in \cref{final-result-theta-BB}. In other  solutions~\cite{Nelson:1983zb,Barr:1984fh}, the finite contributions in \cref{Eq:deltathetaALPresult2Finite} may also be  relevant.} 

\section{Neutron EDM and {cEDM} impact}
\label{sec:impact}
The  stringent experimental limit on the  nEDM, $d_n < 1.8\times 10^{-26}\,e\cdot$cm (90\% C.L.)~\cite{Abel:2020pzs,Pendlebury:2015lrz}, sets important constraints on CP-odd couplings.  Furthermore future experiments are proposed to probe the nEDM at level of $d_n\sim (2-3)\times 10^{-28}\,e\cdot$cm~\cite{Ahmed_2019}. The contribution  of the $\bar{\theta}$-parameter to the nEDM, together with that stemming 
 from the up, down and strange  qEDMs and cEDMs, $d_{u, d,\bel{s}}$ and $\tilde{d}_{u,d,s}$,   can be estimated\footnote{Note that these estimates get modified in the presence of a QCD axion (in addition to the ALP), see \cref{App:Bounds with PQ}.  } to be ~\cite{Kley:2021yhn,Gupta:2018lvp,Pospelov:2005pr,Pospelov:2000bw} 
\begin{align}
d_n&= 0.6(3)\times 10^{-16}\,\bar{\theta}\,[e\cdot\mathrm{cm}]\nn\\
&%\hspace{0.2 cm} 
-0.204(11)\,d_u 
+0.784(28)\,d_d
- 0.0028(17)\,d_s \nonumber \\ 
&%\hspace{1cm}
{- 0.32(15)\,e\,\tilde{d}_u
+0.32(15)\,e\,\tilde{d}_d 
- 0.014(7)\,e\,\tilde{d_s}\,.}
\label{Eq:nEDMthetadudd}
\end{align}
The coefficients of the qEDMs have been obtained with a $\sim5\%$   accuracy by the lattice computation in Ref.~\cite{Gupta:2018lvp},  
whereas the rest of the coefficients, which  present larger errors, are taken from QCD sum rule estimates~\cite{Pospelov:2005pr,Pospelov:2000bw,Hisano:2012cc}.\footnote{The lattice results in Ref.~\cite{Gupta:2018lvp} for the qEDM contributions are well approximated by those obtained with QCD sum rules~\cite{Hisano:2012cc}, once the updated values of the chiral condensates are used, see \cref{App:nEDM estimates from sum rules}.} In particular, we use the analytical estimates computed in Ref.~\cite{Hisano:2012cc} and update some of the input parameters, see \cref{App:nEDM estimates from sum rules} for additional details.  

Even though nEDM experiments provide at the moment the leading bounds on the CP-odd couplings under discussion, storage ring
facilities~\cite{Alexander:2022rmq} are expected to provide limits on the proton EDM (pEDM) of the order of $d_p\sim 10^{-29}\,e\cdot$\,cm in the near future~\cite{Balashov:2007zj}. Therefore, the projections of the pEDM bounds are competitive with those
for future nEDM  measurements.   The corresponding dependence of the pEDM reads~\cite{Cirigliano:2019vfc} 
\begin{align}
d_p&= -1.0(5) \times 10^{-16}\,\bar{\theta}\,[e\cdot\mathrm{cm}]\nn\\
&+ 0.784(28)\,d_u -0.204(11)\,d_d- 0.0028(17)\,d_s \nonumber \\ 
&-0.26(14)\,e\,\tilde{d}_u +0.28(14)\,e\,\tilde{d}_d + 0.02(1)\,e\,\tilde{d}_s\,,
\label{Eq:pEDMthetadudd}
\end{align} 
The estimates for  $d_p(\bth)$ and $d_p(\tilde{d}_{u,d})$ are obtained from the nEDM analytical formulas in Ref.~\cite{Hisano:2012cc} by interchanging $u\leftrightarrow d$, see \cref{App:nEDM estimates from sum rules} for details. The coefficient of ${d}_{u,d}$ in the pEDM formula is taken again from the lattice result in Ref.~\cite{Gupta:2018lvp}. Experimental limits also follow from measurements of atomic EDMs, such as that of ${}^{199}$Hg \cite{PhysRevLett.116.161601}. We do not include them in our analysis as they currently give comparable bounds. 

Using our final result for the ALP contribution to $\bar\theta$ in Eq.~(\ref{final-result-theta-BB}), and barring 
cancellations with the $\bth_0$ value or other terms, the following constraint on the CP-odd ALP parameters involving the top quark is obtained 

\begin{align}
&\left|\frac{\operatorname{Im}[\bld{C}_{Q}^{13}\bld{C}_{u_R}^{*13}]}{f_a^2}\right|_{\bth}\lesssim \Big(\frac{m_t^2}{m_a^2+m_t^2}\Big)\, 2\times 10^{-17}\,\,\text{GeV}^{-2} \,.
\label{eq:C_13_bound}
\end{align}
Here and below we conservatively assume $\log f_a^2/m_a^2 \sim 1$, and also disregard the running of the quark masses (although the bounds could become stronger by over an order of magnitude for $f_a/m_a >10^2$, see \cref{app-running}).  Projections from future nEDM and pEDM measurements  are expected to improve this bound by 3 orders of magnitude. 

\begin{table}[h!]
\renewcommand{\arraystretch}{1.5}
\begin{align*}
\begin{array}[c]{cc}
\hline\\[-10pt]
\left|\operatorname{Im}[\bld{C}_{Q}^{13}\bld{C}_{u_R}^{*13}]/f_a^2\right|\, < 
& \left(\frac{m_t^2}{m_a^2 + m_t^2}\right)
\,2 \times 10^{-17}\\[1mm]
\left|\operatorname{Im}[\bld{C}_{Q}^{23}\bld{C}_{u_R}^{*23}]/f_a^2\right|\, < 
& \left(\frac{m_t^2}{m_a^2 + m_t^2}\right)
\,1 \times 10^{-14}\\[1mm]
\left|\operatorname{Im}[\bld{C}_{Q}^{13}\bld{C}_{d_R}^{*13}]/f_a^2\right|\, < 
& \left(\frac{m_b^2}{m_a^2 + m_b^2}\right)
\,3 \times 10^{-12}\\[1mm]
\left|\operatorname{Im}[\bld{C}_{Q}^{12}\bld{C}_{u_R}^{*12}]/f_a^2\right|\, < 
& \left(\frac{m_c^2}{m_a^2 + m_c^2}\right)
\,5\times 10^{-11}\\[1mm]
\left|\operatorname{Im}[\bld{C}_{Q}^{23}\bld{C}_{d_R}^{*23}]/f_a^2\right|\, < 
& \left(\frac{m_b^2}{m_a^2 + m_b^2}\right)
\,6 \times 10^{-11}\\[1mm]
\left|\operatorname{Im}[\bld{C}_{Q}^{12}\bld{C}_{d_R}^{*12}]/f_a^2\right|\, < 
& \left(\frac{m_s^2}{m_a^2 + m_s^2}\right)
\,3 \times 10^{-7}\\[5pt]
\hline
\end{array}
\end{align*}
\caption{
{\it ALP case.} Bounds on $|\operatorname{Im}[\bld{C}_{Q}^{ij}\bld{C}_{q_R}^{*ij}]/f_a^2|$  in GeV$^{-2}$ obtained from the $\bar{\theta}$ correction. 
}
\label{Tab: bounds for different couplings}
\end{table}

\begin{figure}[h]
    \centering
    \begin{subfigure}{0.49\textwidth}
        \includegraphics[width = \textwidth]{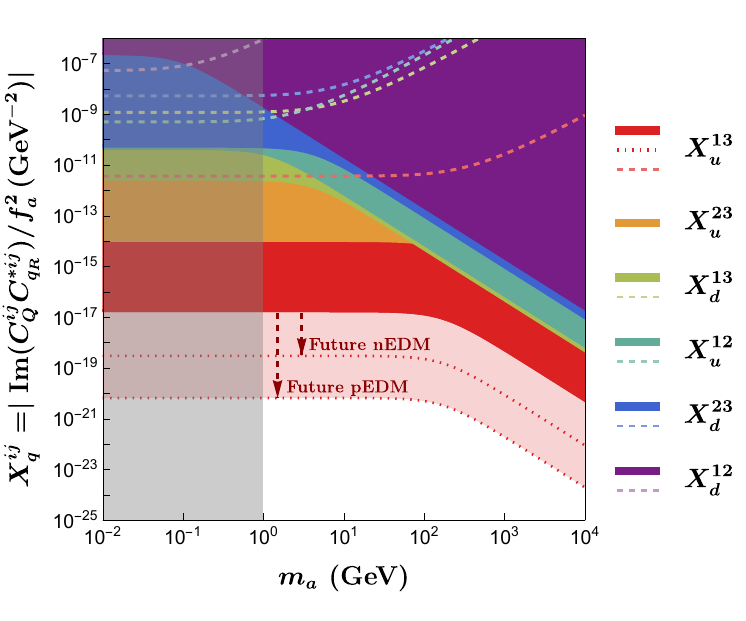}
    \end{subfigure}
    \caption{{\it ALP case.}  Bounds as a function of $m_a$. Upper bounds on $X_{q}^{ij} \equiv \left|\operatorname{Im}\left(\bld{C}_Q^{ij}\bld{C}_{q_R}^{ij}\right)/f_a^2\right|$ from the $\bar{\theta}$ correction (solid regions) and from qEDMs and cEDMs (dashed lines). Future bounds stemming from nEDM and pEDM projections ~\cite{EuropeanStrategyforParticlePhysicsPreparatoryGroup:2019qin} are indicated for $\boldsymbol{X}^{13}_ u$ with a red dotted line. 
    For ALP masses near or below the QCD threshold ($m_a \lesssim 1$\,GeV, shaded region) the results are outside the validity of our approximations, see text. 
    }
    \label{fig:PlotBounds}
\end{figure}

In \cref{Tab: bounds for different couplings}, we show the corresponding results for all the other combinations of couplings $\left|\operatorname{Im}[\bld{C}_{Q}^{ij}\bld{C}_{q_R}^{*ij}]/f_a^2\right|$. {\it All bounds  in this paper consider one coupling combination at a time.} 
We have used the PDG values for the quark masses~\cite{Workman:2022ynf}.
The same bounds are also represented with solid colors in Fig.~\ref{fig:PlotBounds} as a  function of $m_a$.  The region below $m_a=1$ GeV appears shaded in grey --here and in all figures below--  since it lies outside the validity of our approximations and a proper treatment would require an alternative computation where the ALP is included in the chiral Lagrangian, similarly to Refs. \cite{Dekens:2022gha,DiLuzio:2023edk}. 
 For improved plot clarity, future projections are exclusively depicted for the most constrained coupling combination, $\left|\operatorname{Im}\left(\bld{C}_Q^{13}\bld{C}_{q_R}^{13}\right)\right|$. However, it is important to note that all solid regions will be shifted downwards by the same factor based on forthcoming nEDM and pEDM measurements.

\subsection*{Comparison with  quark EDM contributions}

\begin{figure}[h]
    \centering
    \begin{subfigure}{0.49\textwidth}
        \includegraphics[width = \textwidth]{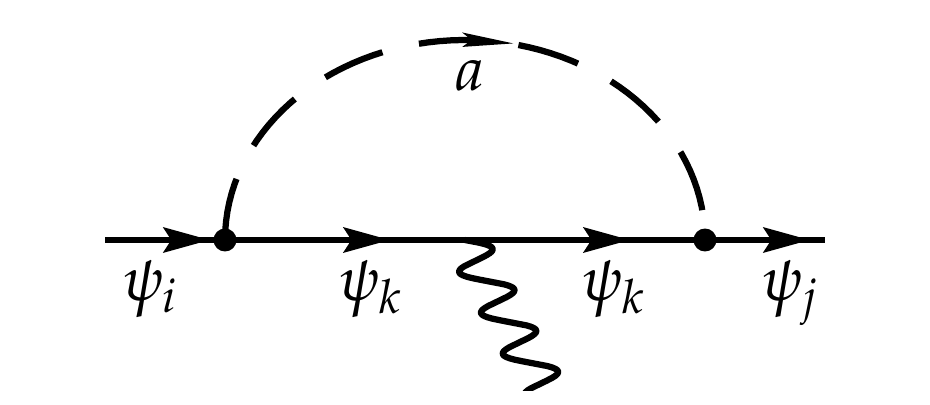}
    \end{subfigure}
    \caption{ALP-induced one loop contribution to the fermion-photon vertex as considered in Ref.\,\cite{DiLuzio:2020oah}. }
    \label{fig:1-loop-diagsupEDM}
\end{figure}

 The Lagrangian in \cref{Eq:Initial Lagrangian} also generates a  contribution to the individual qEDMs  $d_{u,d,s}$ and cEDMs $\tilde{d}_{u,d,s}$, which in turn contribute   to the final nEDM value through Eq.~(\ref{Eq:nEDMthetadudd}).  The corresponding Feynman diagram -- see  Fig.~\ref{fig:1-loop-diagsupEDM} -- has been previously  computed~\cite{DiLuzio:2020oah,DasBakshi:2023lca}, 
and for the up-quark EDM it contributes as 
\begin{align}
\frac{d_{u}}{e} &\equiv \frac{Q_u}{32\pi^2 f_a^2} \operatorname{Im}\left[\mathbf{C}_Q \boldsymbol{G}\,\mathbf{C}_{u_R}^\dagger \right]_{11}\,,
\label{Eq:uEDMcLcR}
\end{align}
 with $\boldsymbol{G} = \operatorname{diag}\left(G(x_u),\,G(x_c),\,G(x_t)\right)$ and
\begin{align}
  G(x_k) &\equiv m_a\,x_{k}^{3/2}\frac{(3- 4\,x_{k} + x_{k}^2 + 2\,\log(x_{k}))}{(x_{k}-1)^3}\nn\\ 
  &\sim \frac{m_k}{m_a^2}\texttt{min}(m_a^2,3m_k^2)\,,
  \label{eq:fermion_EDM}
\end{align}
where $x_{k}\equiv m_{k}^2/m_a^2$\,, $Q_u = +2/3$ is the charge of the up-quark in units of $e$, and the last expression captures the order of magnitude for either  $m_a \ll m_k$ or $m_a \gg m_k$. An analogous expression holds for the ALP induced $d_d$.  Also, the same Feynman diagram depicts the one-loop contribution to the quark cEDM $\tilde{d}_{q}$ by replacing the photon with a gluon,  and its computation can be simply recast from that for the  qEDM as $\tilde{d}_{q} = d_{q}/e\,Q_{u,d}$, where the $g_s$ factor is customarily factored out in the definition, see Eq.~(\ref{Eq:QCD Low energy Lag}).

 The ${d}_{q}$ and $\tilde{d}_{q}$ contributions to the nEDM are comparable, and again the strongest bound corresponds to that involving the top quark in the loop, 
\begin{equation}
\left|\frac{\operatorname{Im}(\bld{C}_{Q}^{13}{\bld{C}_{u_R}^{*13}})}{f_a^2}\right|_{{d}_{q}
,\,\tilde{d}_{q}}
 \lesssim 3.7 \times 10^{-12}\,\text{GeV}^{-2}\,,
 \label{eq:bound_diLuzio}
\end{equation}
which is several orders of magnitude weaker than that obtained in \cref{eq:C_13_bound} from  $ \bth$.

\begin{table}[h!]
\renewcommand{\arraystretch}{1.5}
\begin{align*}
\begin{array}[c]{ccc}
\multirow{2}{*}[1.1mm]{\text{Combination}}& \text{$\bar{\theta}$-bounds}& \multirow{1}{*}{\text{qEDM \& cEDM}}\\[-2mm]
&  \text{{\small(GeV$^{-2}$)}} & \text{{\small(GeV$^{-2}$)}}    \\[0mm]
\hline\\[-10pt]
\left|\operatorname{Im}[\bld{C}_{Q}^{13}\bld{C}_{u_R}^{*13}]/f_a^2 \right|
& 
1.8 \times 10^{-17}
& 3.7\times 10^{-12}
\\[1mm]
\left|\operatorname{Im}[\bld{C}_{Q}^{23}\bld{C}_{u_R}^{*23}]/f_a^2\right|
& 
1.1 \times 10^{-14}
& 8.3\times 10^{-8}
\\[1mm]
\left|\operatorname{Im}[\bld{C}_{Q}^{13}\bld{C}_{d_R}^{*13}]/f_a^2\right|
& 
1.1 \times 10^{-12}
& 1.9 \times 10^{-9}
\\[1mm]
\left|\operatorname{Im}[\bld{C}_{Q}^{12}\bld{C}_{u_R}^{*12}]/f_a^2\right|
& 
2.7 \times 10^{-12}
& 2.3 \times 10^{-9}
\\[1mm]
\left|\operatorname{Im}[\bld{C}_{Q}^{23}\bld{C}_{d_R}^{*23}]/f_a^2\right|
& 
2.3 \times 10^{-11}
& 8.7 \times 10^{-9}
\\[1mm]
\left|\operatorname{Im}[\bld{C}_{Q}^{12}\bld{C}_{d_R}^{*12}]/f_a^2\right|
& 
8.6 \times 10^{-11}
& 1.2 \times 10^{-5}
\\[5pt]
\hline
\end{array}
\end{align*}
\caption{{\it ALP case.} Bounds from $\bth$ contributions versus qEDM and cEDM bounds.  
Experimental bounds on $\left|\operatorname{Im}[\bld{C}_{Q}^{ij}\bld{C}_{q_R}^{*ij}]/f_a^2\right|$ for the particular case  $m_a = 5$\,GeV. 
 The last column corresponds to the sum of the  qEDM and cEDM contributions. The bounds can be extrapolated to other values of $m_a$ using Eq.~(\ref{Eq:deltathetaALPresult2}) and Eq.~(\ref{eq:fermion_EDM}), respectively.
}
\label{Tab: comparison with EDM + cEDM}
\end{table}  

{Similarly,  separate bounds on the other possible combinations of parameters can be obtained
using \cref{Eq:nEDMthetadudd,eq:fermion_EDM}. As an example, in \cref{Tab: comparison with EDM + cEDM} we compare the bounds for the CP-odd combinations $X_{q}^{ij} \equiv \left|\operatorname{Im}(\bld{C}_Q^{ij}\bld{C}_{q_R}^{ij})/f_a^2\right|$ stemming
 from the qEDM and cEDM with those resulting from our analysis of the $\bth$-parameter,  for an ALP mass $m_a=5$ GeV. Indirect contributions from the charm and bottom (c)qEDM, not included in \cref{Eq:nEDMthetadudd}, give bounds on $\left|\operatorname{Im}[\bld{C}_{Q}^{23}\bld{C}_{u_R}^{*23}]/f_a^2\right|$ \cite{Gisbert:2019ftm,Ema:2022pmo}
\cref{fig:PlotBounds} depicts,   as a function of $m_a$, the bounds from $\bth$ in solid regions and those from qEDM and  cEDM  with dashed lines.  It is seen that, for $m_a\gtrsim 1$\,GeV,  the dominant bounds on $X^{ij}_q$ arise from the ALP contribution to $\bar{\theta}$. 
% Notice as well that the  qEDM and cEDM   contributions do not allow to extract a bound on the combination $\operatorname{Im}[\bld{C}_{Q}^{23}\bld{C}_{u_R}^{*23}]/f_a^2$,    while the $\bar{\theta}$  parameter is sensitive to it and leads to a strong bound (in blue).
% \pablo{A bound for the combination $\operatorname{Im}[\bld{C}_{Q}^{23}\bld{C}_{u_R}^{*23}]/f_a^2$ follows from qEDM and cEDM contributions follow}
% Notice as well that the  qEDM and cEDM   contributions do not allow to extract a bound on the combination $\operatorname{Im}[\bld{C}_{Q}^{23}\bld{C}_{u_R}^{*23}]/f_a^2$,    while the $\bar{\theta}$  parameter is sensitive to it and leads to a strong bound (in blue).

Complementary bounds on these parameters can be obtained from CP violation  in $\Delta F=2$ processes \footnote{We thank Andreas Trautner and the referee for raising this issue.}. For example, from $K-\bar K$ mixing one obtains \cite{UTfit:2007eik}
 % \begin{align}
 %  \frac{1}{\Lambda^2} \big|\operatorname{Im}\big[(K_{12}+K_{21}^{*})^2\big]\big|\lesssim 10^{-20} \left(\frac{m_a}{5 \,\text{GeV}}\right)^2 \text{GeV}^{-2} 
 %  \end{align} 
  \begin{align}
  \frac{1}{f_a^2} \Big|\operatorname{Im}\big[(C_Q^{12}-C_{d_R}^{12})^2\big]\Big|\lesssim 10^{-13} \left(\frac{m_a}{5 \,\text{GeV}}\right)^2 \text{GeV}^{-2} 
  \label{Eq:KKbarmixing}
  \end{align}
This strong bound corresponds to a different direction in parameter space, and has a different parametric dependence on $m_a$.}

We have cross-checked our results by performing the computation in an alternative basis, see  \cref{App:ChiralityFlip}. Indeed, the
chirality-preserving ALP-fermion interactions in \cref{Eq:Initial  Lagrangian} can be traded for a specific combination of  Yukawa-like
operators in the so-called ``chirality-flip basis'' via a suitable field redefinition of the fields. 
  As this is only a change of basis, the number of free parameters does not change.

\vspace{0.5cm}

\section{Generic scalar with CP-odd fermionic interactions} 
\label{sec:generic_scalar_with_fermionic_interactions} 
 \begin{figure}
    \centering
    \begin{subfigure}{0.45\textwidth}
        \includegraphics[width = \textwidth]{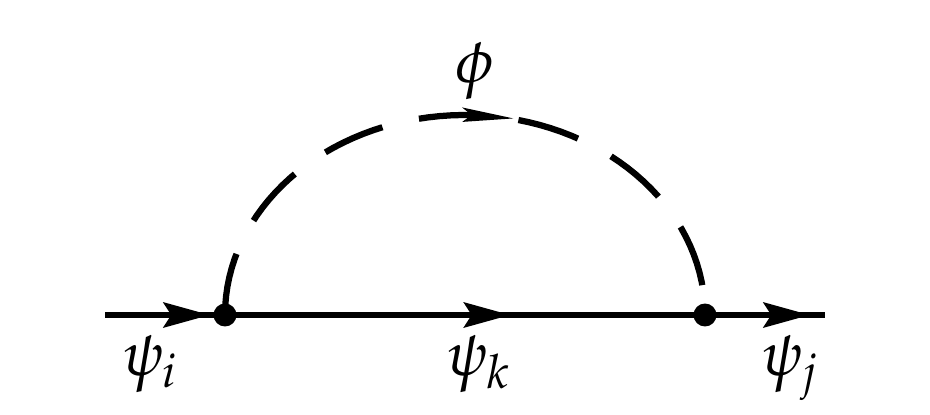}
    \end{subfigure}\\[0.3cm]
    \begin{subfigure}{0.1\textwidth}
        \centering
        \includegraphics[width = 0.7\textwidth]{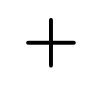}
    \end{subfigure}\\
    \begin{subfigure}{0.4\textwidth}
        \includegraphics[width = \textwidth]{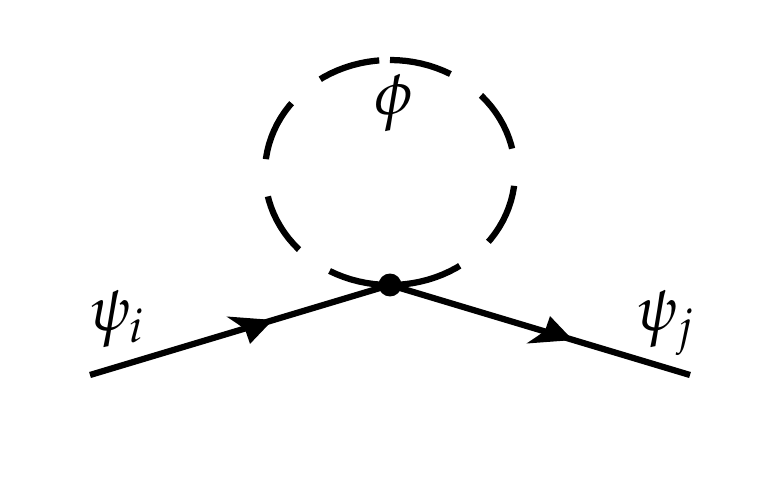}
    \end{subfigure}
    \caption{ Scalar exchange one-loop contribution to the quark masses.}
    \label{fig:1-loop-diags-CF}
\end{figure}

In this section, the previous analysis is extended to a theory of a singlet scalar $\phi$ with mass $m_\phi$ and  coupled to SM fermions without imposing shift symmetry.

  The most general CP-odd fermion-scalar interactions are described by a Lagrangian exhibiting Yukawa-like left-right interactions,
  \begin{align}
 \mathcal{L}&\supset \bar{u}_L\,v\,\left[i\,\boldsymbol{K}_u
   \frac{\phi}{\Lambda}\
                      +\boldsymbol{F}_u\frac{\phi^2}{\Lambda^2}\right] u_R\nonumber\\
    &\hspace{1cm}+\bar{d}_L\,v\,\left[i
   \frac{\phi}{\Lambda}\boldsymbol{K}_d
                      +\frac{\phi^2}{\Lambda^2}\boldsymbol{F}_d\right] d_R+\text{h.c.}
  \label{eq:chirality-flip-phi}
 \end{align}
where  $\boldsymbol{K}_q$ and $\boldsymbol{F}_q$ are arbitrary $3\times3$ dimensionless coefficient matrices in flavour space
 and $\Lambda$ denotes the new physics scale. The  dependence on the Higgs vacuum expectation value, $v = 246$\,GeV, is a remnant of the couplings ancestry above the electroweak scale, which corresponds to Yukawa-like operators of mass dimension 5 ($\boldsymbol{K}_q$ terms) and 6 ($\boldsymbol{F}_q$ terms).  The $\mathcal{O}(1/\Lambda^2$)  $\boldsymbol{F}_q$ terms 
 have been included as this is required by the consistency of the EFT analysis at one-loop.  They contribute to $\bth$ through the lower diagram in \cref{fig:1-loop-diags-CF},  at the same order as the first one (which presents two insertions of the coupling  proportional to $\boldsymbol{K}_d/\Lambda$).  Even considering only the $\boldsymbol{K}_q$ terms, this Lagrangian has more free parameters than that for the ALP in Eq.~(\ref{Eq:Initial Lagrangian})  (see also App.~\ref{App:ChiralityFlip}).  Those extra parameters include flavour-diagonal ones which can be complex, thus sourcing additional CP-violation effects, in contrast to the case of the ALP theory.

In the literature, it is also common to use the alternative notation
\begin{align}
\bar{q}_L\,i\,v\,\frac{a}{f_a} \bld{K}_q\,q_R \equiv \frac{a}{f_a}\,v\,\left(\bar{q}\,\bld{y}_{q S}\,q + i\,\bar{q}\,\bld{y}_{q P}\gamma_5q\,\right)\,,
\label{Eq:literature}
 \end{align}
with
\begin{align}
  \bld{y}_{q S} &\equiv i\,\frac{\bld{K}_q-\bld{K}_q^\dagger}{2}\,, &
  \bld{y}_{q P} &\equiv \frac{\bld{K}_q+\bld{K}_q^\dagger}{2}\,.
  \label{Eq:literature2}
\end{align}

The contribution of  the $\boldsymbol{K}_q$ and $\boldsymbol{F}_q$ parameters to $\bth$ through its running reads: 
\begin{widetext}
\begin{align}
&\bar{\theta}\left(\mu_{IR}\right)\simeq \bth_0+\frac{v^2}{16\pi^2\Lambda^2 }\times 
\Bigg(\sum_{i,k} \bigg[\frac{m_{u_k} \operatorname{Im}\left(\boldsymbol{K}_u^{ik}\boldsymbol{K}_u^{ki}\right)}{m_{u_i}}-\frac{m_{\phi}^2\operatorname{Im}\left(\boldsymbol{F}_u^{ik}\right)}{v\,m_{u_i}}\bigg] \log \frac{\Lambda^2}{\texttt{max}(m_{\phi}^2,m_{u_k}^2)}\nn \\
&\hspace{6 cm}+\sum_{i,k} \bigg[\frac{m_{d_k} \operatorname{Im}\left(\boldsymbol{K}_d^{ik}\boldsymbol{K}_d^{ki}\right)}{m_{d_i}}-\frac{m_{\phi}^2\operatorname{Im}\left(\boldsymbol{F}_d^{ik}\right)}{v\,m_{d_i}}\bigg]\log \frac{\Lambda^2}{\texttt{max}(m_{\phi}^2,m_{d_k}^2)}\Bigg) \,. 
\label{eq:delta_theta_flipping}
\end{align}
\end{widetext}
 In turn, their contribution to the up-quark EDM is given by 
\begin{align}
\frac{d_{u}}{e} &\equiv \frac{Q_u}{32\pi^2}\frac{v^2}{ \Lambda^2} \operatorname{Im}\left[\mathbf{K}_q \boldsymbol{G'}\,\mathbf{K}_q^\dagger \right]_{11}\,,
\label{Eq:uEDMKK}
\end{align}
with $\boldsymbol{G}' = \operatorname{diag}\left(G'(x_u),\,G'(x_c),\,G'(x_t)\right)$,  where $x_k\equiv m_k^2/m^2_\phi$, and
\begin{align}
  G'(x_k) &\equiv \frac{x_{k}^{1/2}}{m_{\phi}}\frac{(3- 4\,x_{k} + x_{k}^2 + 2\,\log(x_{k}))}{(x_{k}-1)^3}\,,\nn
  \label{eq:fermion_EDM_KK}
\end{align}
and analogously for the other flavours and for the quark cEDMs \cite{DiLuzio:2020oah}.

\begin{figure}[h!]
    \centering
    \begin{subfigure}{0.49\textwidth}
        \includegraphics[width = \textwidth]{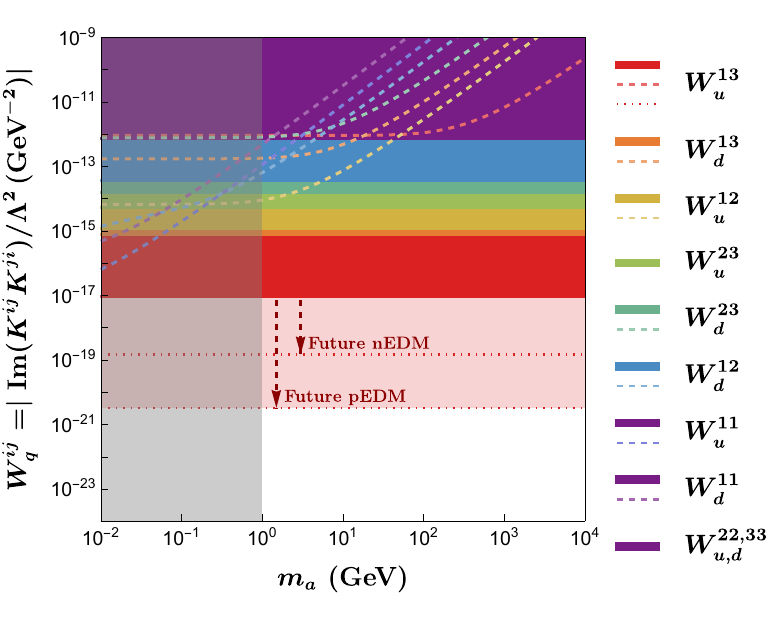}
    \end{subfigure}
    \caption{{\it General scalar.} Upper bounds on $W_{q}^{ij} \equiv \left|\operatorname{Im}(\bld{K}_q^{ij}\bld{K}_{q}^{ji})/\Lambda^2\right|$ stemming from the contributions of $\bar{\theta}$  (solid regions) and  from the sum of qEDMs and cEDMs (dashed lines) to the nEDM. The red dotted line shows the projected bounds on $W_u^{13}$ from  future  nEDM and pEDM experiments ~\cite{Alexander:2022rmq,Balashov:2007zj}. The grey shaded area is as described in Fig.~\ref{fig:PlotBounds}.}   
 \label{fig:PlotBoundsKK}
\end{figure}

\begin{figure}[h!]
    \centering
    \begin{subfigure}{0.45\textwidth}
        \includegraphics[width = \textwidth]{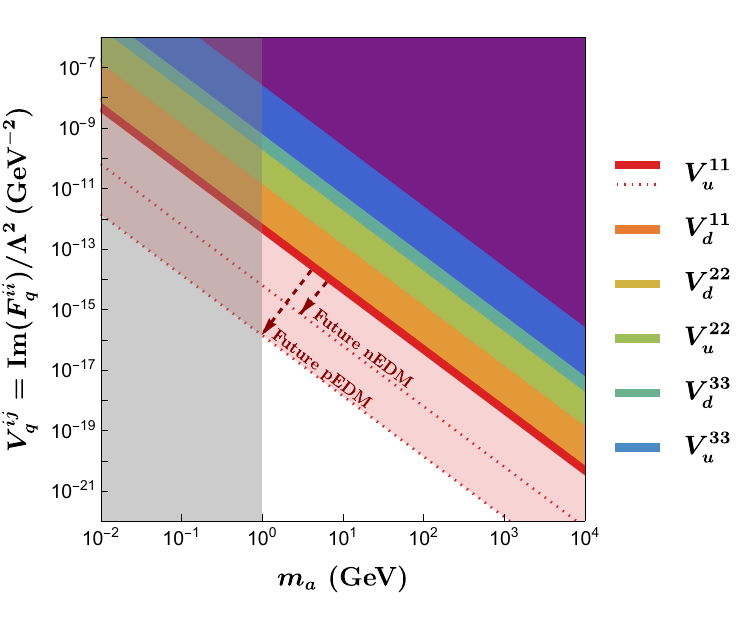}
    \end{subfigure} 
    \caption{{\it General scalar.} Upper bounds on $V_{q}^{ij} \equiv \left|\operatorname{Im}\left(F_q^{ij}\right)/\Lambda^2\right|$ stemming from the contributions of $\bar{\theta}$ (solid regions) to the nEDM.
 The red dotted line shows the projected bounds on $V_u^{11}$ from  future  nEDM and pEDM experiments~\cite{Alexander:2022rmq,Balashov:2007zj}. The grey shaded area is as described in previous plots.} 
    \label{fig:PlotBoundsF}
\end{figure}

Stringent restrictions on this enlarged $\{\bld{K}_q,\bld{F}_q \}$ parameter space  follow  from the nEDM experimental limit~\cite{Abel:2020pzs,Afach:2015sja}, with an impact comparable to that found in the previous sections for the CP-odd couplings of a generic ALP. Using again \cref{Eq:nEDMthetadudd}, the nEDM is seen to be sensitive to the combinations  $W_q^{ij} \equiv \left|\operatorname{Im}(\bld{K}_{q}^{ij} \bld{K}_{q}^{ji})/\Lambda^2\right|$ and $V_q^{ii} \equiv \left|\operatorname{Im}(\bld{F}_q^{ii})/\Lambda^2\right|$. The corresponding bounds are displayed  in \cref{tab:KK-bounds}  and depicted as a function of $m_\phi$ in  Figs.~\ref{fig:PlotBoundsKK} and \ref{fig:PlotBoundsF}. For most   $W_q^{ij}$'s, the constraint imposed by their contribution to $\bar{\theta}$ dominates strongly over that from  contributions to qEDMs and cEDMs. In addition, the contributions to $\bar{\theta}$  also provide a handle into several combinations of parameters {---$W_{u}^{33}$ and the $V^{ii}_q$'s ---} 
 which do not contribute to neither the qEDMs nor the cEDMs.
 
Analogously to the ALP case in \cref{Eq:KKbarmixing}, complementary bounds for different directions in the parameter space of the general scalar can be obtained from CP violation  in $\Delta F=2$ processes such as meson mixing \cite{UTfit:2007eik}.
 % \pablo{Complementary bounds on these parameters can be obtained from CP violation  in $\Delta F=2$ processes \footnote{\red{We thank Andreas Trautner for raising this issue. Similar bounds would follow for the case of an ALP.}}. For example, from $K-\bar K$ mixing one obtains \cite{UTfit:2007eik}
%\begin{align}
%\frac{1}{\Lambda^2} \{\operatorname{Im}(K_{12}^2), \operatorname{Im}(K_{12}^2), \operatorname{Im}(K_{12}K^{*}_{12})\}\lesssim 10^{-20} \left(\frac{m_a}{5 \,\text{GeV}}\right)^2 \text{GeV}^{-2} 
%\frac{1}{\Lambda^2} \big|\operatorname{Im}\big[(K_{12}+K_{21}^{*})^2\big]\big|\lesssim 10^{-20} \left(\frac{m_a}{5 \,\text{GeV}}\right)^2 \text{GeV}^{-2} 
%\end{align} 
% This strong bound corresponds to a different direction in parameter space, and has a different parametric dependence on $m_a$.}

In summary, much as in the case of ALPS,   the  bounds obtained for a general scalar are several orders of magnitude stronger than those existing in the literature, due to the impact of the $\bth$ parameter. Furthermore, some new combinations of parameters have become accessible via the present $\bar{\theta}$ analysis.

\begin{table}
\renewcommand{\arraystretch}{1.5}
\begin{align*}
\begin{array}[c]{ccc}
\multirow{2}{*}[1.1mm]{\text{Combination}}& \text{$\bar{\theta}$-bounds}& \multirow{1}{*}{\text{qEDM \& cEDM}}\\[-2mm]
&  \text{{\small(GeV$^{-2}$)}} & \text{{\small(GeV$^{-2}$)}}    \\[0mm]
\hline\\[-10pt]
\left|\operatorname{Im}(\bld{K}_{u}^{13} \bld{K}_{u}^{31})/\Lambda^2\right|
& 8.9\times 10^{-18}
& 9.0\times 10^{-13}
\\[1mm]
\left|\operatorname{Im}(\bld{K}_{d}^{13} \bld{K}_{d}^{31})/\Lambda^2\right|
& 7.9\times 10^{-16}
& 2.8\times 10^{-13}
\\[1mm]
\left|\operatorname{Im}(\bld{K}_{u}^{12} \bld{K}_{u}^{21})/\Lambda^2\right|
& 1.2\times 10^{-15}
& 3.1\times 10^{-14}
\\[1mm]
\left|\operatorname{Im}(\bld{K}_{u}^{23} \bld{K}_{u}^{32})/\Lambda^2\right|
& 5.2\times 10^{-15}
& 2.1\times 10^{-8}
\\[1mm]
\left|\operatorname{Im}(\bld{K}_{d}^{23} \bld{K}_{d}^{32})/\Lambda^2\right|
& 1.6 \times 10^{-14}
& 1.3\times 10^{-12}
\\[1mm]
\left|\operatorname{Im}(\bld{K}_{d}^{12} \bld{K}_{d}^{21})/\Lambda^2\right|
& 3.5\times 10^{-14}
& 8.2\times 10^{-13}
\\[1mm]
\left|\operatorname{Im}(\bld{K}_{u}^{11} \bld{K}_{u}^{11})/\Lambda^2\right|
& 7.1\times 10^{-13}
& 2.2\times 10^{-12}
\\[1mm]
\left|\operatorname{Im}(\bld{K}_{d}^{11} \bld{K}_{d}^{11})/\Lambda^2\right|
& 7.1\times 10^{-13}
& 8.7\times 10^{-12}
\\[1mm]
\left|\operatorname{Im}(\bld{K}_{d}^{22} \bld{K}_{d}^{22})/\Lambda^2\right|
& 7.1\times 10^{-13}
& 3.8\times 10^{-12}
\\[1mm]
\left|\operatorname{Im}(\bld{K}_{u}^{22} \bld{K}_{u}^{22})/\Lambda^2\right|
& 7.1\times 10^{-13}
& 7.0\times 10^{-10}
\\[1mm]
\left|\operatorname{Im}(\bld{K}_{u}^{33} \bld{K}_{u}^{33})/\Lambda^2\right|
& 7.1\times 10^{-13}
& -
\\[1mm]
\left|\operatorname{Im}(\bld{K}_{d}^{33} \bld{K}_{d}^{33})/\Lambda^2\right|
& 7.1\times 10^{-13}
& 5.5\times 10^{-8}
\\[1mm]
\left|\operatorname{Im}(\bld{F}_{u}^{11})/\Lambda^2\right|
& 1.5\times 10^{-14}
& -
\\[1mm]
\left|\operatorname{Im}(\bld{F}_{d}^{11})/\Lambda^2\right|
& 3.3\times 10^{-14}
& -
\\[1mm]
\left|\operatorname{Im}(\bld{F}_{d}^{22})/\Lambda^2\right|
& 6.5\times 10^{-13}
& -
\\[1mm]
\left|\operatorname{Im}(\bld{F}_{u}^{22})/\Lambda^2\right|
& 8.9\times 10^{-12}
& -
\\[1mm]
\left|\operatorname{Im}(\bld{F}_{d}^{33})/\Lambda^2\right|
& 2.9 \times 10^{-11}
& -
\\[1mm]
\left|\operatorname{Im}(\bld{F}_{u}^{33})/\Lambda^2\right|
& 1.2\times 10^{-9}
& -
\\[2mm]
\hline
\end{array}
\end{align*}
\caption{
{\it General scalar}. Bounds on $\left|\operatorname{Im}(\bld{K}_{q}^{ij} \bld{K}_{q}^{ji})/\Lambda^2\right|$ and $\left|\operatorname{Im}(\bld{F}_{q}^{ii})/\Lambda^2\right|$ stemming from their contribution to the nEDM.  The  bounds on  $\left|\operatorname{Im}(\bld{K}_{q}^{ij} \bld{K}_{q}^{ji})/\Lambda^2\right|$ inferred from their contribution to $\bar{\theta}$  are independent of $m_\phi$, while those stemming from  the sum of qEDMs and cEDMs   are evaluated for $m_\phi = 5$\,GeV. The bounds that stem from the indirect effect of the charm and bottom (c)qEDMs are derived from~\cite{Ema:2022pmo}.
}
\label{tab:KK-bounds}
\end{table}

\section{Discussion}

In this work, we have computed the one-loop contribution to the $\bar \theta$-parameter of an ALP with CP-odd derivative couplings to fermions.  Its impact on the nEDM is such that the bounds obtained on the CP-odd ALP-fermion couplings  are many orders of magnitude stronger than those stemming from the  contributions to up- and down-quark electric and chromo-electric dipole moments. The same conclusion applies to the CP-odd fermionic couplings of a generic scalar particle.

Finally, it is worth mentioning that the strong novel  bounds found above, either for an ALP theory or for a general scalar, do not apply if those theories are extended by including in the Lagrangian --in addition to the ALP-- an extra true QCD axion which solves the strong CP problem via a Peccei-Quinn mechanism. In that case, the QCD axion would absorb all contributions to $\bth$ computed in this work. Still a non-zero induced $\bth_{\text{ind}}$~\cite{Bigi:1991rh,Pospelov:1999ha,Pospelov:1997uv,Pospelov:2000bw} would remain,  leading to weaker but significant bounds on the studied couplings \cite{Dekens:2022gha,DiLuzio:2020oah}. We expand on that case in \cref{App:Bounds with PQ}.

\vspace{1cm}

% \red{{ \bf Note added in proof.} Shortly before making public our results, Ref.~\cite{Banno:2023yrd} appeared where the scalar contributions to $\bar{\theta}$ is discussed within a very different approach. Their  one-loop analysis differs from ours  in that :  i)  the case of an ALP is not discussed, but just that of a generic scalar; ii) the Lagrangian in that reference is not $SU(2)\times U(1)$ gauge invariant and thus it misses altogether the need to include $\mathcal{O} (1/\Lambda^2)$ operators for consistency of the EFT; 
% % (in fact their contribution to thetabar would correspond to the K-dependent part of our Eq. 22, if the former was weighted down by the v^2/Lambda^2 factor required by gauge invariance);  i
%  ii)  no phenomenological bounds are derived. }

\vspace{1cm}

\paragraph*{
{\bf Acknowledgments.}}---%
We thank Jes\'us Bonilla and Mar\'ia Ramos for illuminating discussions in the early stages of this project. We thank Jonathan Kley for pointing us to the most updated estimation of the qEDM contributions to the nEDM. We also thank Luca di Luzio and Maxim Pospelov for clarifications on their work, as well as  Ilaria Brivio, Zoltan Ligeti, Alessandro Valenti and Andreas Trautner for illuminating comments and advice. This project has received funding /support from the European Union's Horizon 2020 research and innovation programme under the Marie Sklodowska-Curie grant agreement No.~860881-HIDDeN, and under the Marie Sklodowska-Curie Staff Exchange  grant agreement No.~101086085-ASYMMETRY. The work of B.Gr.~and P.Q.\ is supported in part by the U.S.~Department of Energy Grant No. DE-SC0009919. The work of V.E. was supported by the Spanish MICIU
through the National Program FPI-Severo Ochoa (grant number PRE2020-094281). B.Ga. and V. E. acknowledge as well partial financial support from the Spanish Research Agency (Agencia Estatal de Investigaci\'on) through the grant IFT Centro de Excelencia Severo Ochoa No CEX2020-001007-S, the grants PID2019-108892RB-I00 and PID2022-137127NB-I00 funded by MCIN/AEI/10.13039/501100011033/ FEDER, UE. B.Ga. and V. E. thank very much the Particle Physics group of the University of California, San Diego, where part of this work was carried out.\\[5pt]

\appendix

\section{Chirality-flip basis}
\label{App:ChiralityFlip}

As a double check and to facilitate the comparison of our results with the rest of the literature, we have also computed the ALP-fermion  contribution to  $\bth$ in the chirality-flip basis. It is well-known~\cite{Georgi:1986df,Chala:2020wvs,Bonilla:2021ufe} that by applying the transformation
\begin{align}
u_{L} &\longrightarrow e^{i \frac{a}{f_a}\boldsymbol{C}_{Q}}
          u_{L} \,, & d_{L} &\longrightarrow e^{i \frac{a}{f_a}\boldsymbol{C}_{Q}}
          d_{L} \,,\\
  u_{R} &\longrightarrow e^{i \frac{a}{f_a}\boldsymbol{C}_{u_R}}
          u_{R}\,,  &  d_{R} &\longrightarrow e^{i \frac{a}{f_a}\boldsymbol{C}_{d_R}}
          d_{R} \, 
\end{align}
to the Lagrangian in \cref{Eq:Initial Lagrangian} one can trade the derivative ALP-fermionic interactions by Yukawa-like left-right interactions as described by
 \begin{align}
 \mathcal{L}&\supset \bar{u}_L\,v\,\left[i
   \frac{a}{f_a}\boldsymbol{K}_u
                      +\frac{a^2}{f_a^2}\boldsymbol{F}_u\right] u_R\nonumber\\
    &\hspace{1cm}+\bar{d}_L\,v\,\left[i
   \frac{a}{f_a}\boldsymbol{K}_d
                      +\frac{a^2}{f_a^2}\boldsymbol{F}_d\right] d_R+\text{h.c.} +\dots
  \label{eq:chirality-flip}
 \end{align} 
where the $\boldsymbol{K}_q$ and $\boldsymbol{F}_q$ coefficient matrices  are  given by~  \begin{align}
\begin{aligned}
v\,\boldsymbol{K}_q & \equiv \boldsymbol{C}_Q \boldsymbol{M}_q-\boldsymbol{M}_q \boldsymbol{C}_{q_R} \,,\\
2\,v\,\boldsymbol{F}_q & \equiv 2 \boldsymbol{C}_Q \boldsymbol{M}_q \boldsymbol{C}_{q_R}-\boldsymbol{C}_Q^2 \boldsymbol{M}_q-\boldsymbol{M}_q \boldsymbol{C}_{q_R}^2\,,
\label{Eq:DerivativeChiralityFlipCorrepondance}
\end{aligned}
\end{align}
and the ellipsis stands for higher order terms $\mathcal{O}(1/f_a^3)$.

 The Lagrangian in Eq.~(\ref{eq:chirality-flip})  is formally the same than that in Eq.~(\ref{eq:chirality-flip-phi}) for a general scalar (with the replacement $\phi\to a$ and $\Lambda \to f_a$), but the relations in Eq.~(\ref{Eq:DerivativeChiralityFlipCorrepondance}) reduce its degrees of freedom. They ensure that  it is  shift-symmetric under $a \to a + \rm{constant}$ and equivalent to the ALP Lagrangian in Eq.(\ref{Eq:Initial Lagrangian}).

In this chirality-flip basis the  two diagrams in  \cref{fig:1-loop-diags-CF}  contribute to the one-loop corrections to the quark masses (in contrast with the derivative basis where only the first topology contributes). In particular, it is this second diagram that exhibits the quadratic divergence $\propto m_a^2$. It is straightforward to check that the result obtained in the derivative basis, \cref{final-result-theta-BB}, is recovered by substituting \cref{Eq:DerivativeChiralityFlipCorrepondance} into \cref{eq:delta_theta_flipping}. 

\section{nEDM estimates from QCD sum rules without a Peccei-Quinn symmetry}
\label{App:nEDM estimates from sum rules}
Various methods, including  naive dimensional analysis~\cite{Manohar:1983md}, chiral lagrangians~\cite{Crewther:1979pi,Guo:2012vf}, lattice simulations~\cite{Gupta:2018lvp} and QCD sum rules~\cite{Shifman:1978bx,Pospelov:2005pr} can be employed to obtain estimates of the nEDM and pEDM as a function of $\bar{\theta},\, d_{q}$ and $\tilde{d}_{q}$. Their corresponding results do not always agree with each other and present a large span of uncertainties.

Thoughout this paper we will use the lattice result in Ref.~\cite{Gupta:2018lvp} for the coefficients of the qEDMs  in the nEDM and pEDM formulae, 
\begin{align}
d_{n}( d_{q})&=-0.204(11)\,d_u 
+0.784(28)\,d_d
- 0.0028(17)\,d_s\nn\\
d_{p}( d_{q})&=+ 0.784(28)\,d_u -0.204(11)\,d_d- 0.0028(17)\,d_s
\label{Eq:Latice nums qEDM}
\end{align}
Note that the estimation in Ref.~\cite{DiLuzio:2020oah} has a typographical error as the numerical coefficients for 
 $d_u$ and $d_d$  
 appear interchanged with respect to the lattice results in Ref.~\cite{Gupta:2018lvp}. 

For the rest of the coefficients  $d_{n}(\bar{\theta}, \tilde{d}_q)$ we will use the analytical formulas from QCD sum rule estimates in Ref.~\cite{Hisano:2012cc} for the nEDM. The formula for $d_{p}(\bar{\theta}, \tilde{d}_q)$ is obtained by  interchanging $u\leftrightarrow d$ in $d_{n}(\bar{\theta}, \tilde{d}_q)$. The result reads
\begin{align}
d_{n,p}\left(\bar{\theta}, \tilde{d}_q, d_q\right)=-c_0 \frac{\langle\bar{q} q\rangle}{\lambda_N^2m_{n,p}} {\Theta_{n,p}}\left(\bar{\theta}, \tilde{d}_q, d_q\right), \label{Eq:AnalyticsnpEDMHisano1}
\end{align}
where $c_0=1.8 \times 10^{-2}\,{\rm GeV}^4$~\cite{Hisano:2012cc}, $m_{n,p}$ respectively denote the neutron and proton mass,   the value of the chiral condensate  will be taken from the recent computation in Ref.~\cite{Gubler:2018ctz} $\langle\bar{q} q\rangle=-(0.272(5) \mathrm{GeV})^3$ and the low-energy constant $\lambda_N=-0.0436(131) \mathrm{GeV}^3$ will be taken from a lattice calculation \cite{Hisano:2012cc}. The  ${\Theta_{n,p}}$ functions are given by  
\begin{widetext}
\fontsize{9.5pt}{11pt}\selectfont
\begin{align}
\Theta_n\left(\bar{\theta}, \tilde{d}_q, d_q\right)&=  \chi m_*\left[\left(4 e_d-e_u\right)\left(\bar{\theta}-\frac{m_0^2}{2} \frac{\tilde{d}_s}{m_s}\right)+\frac{m_0^2}{2}\left(\tilde{d}_d-\tilde{d}_u\right)\left(\frac{4 e_d}{m_u}+\frac{e_u}{m_d}\right)+  \frac{m_0^2}{2}\,\left(\frac{4\,e_d}{m_s}\tilde{d}_d-\frac{e_u}{m_s}\tilde{d}_u\right)\right]+\nn\\
 &\hspace{9cm}+\big(\kappa-\frac{1}{2}\xi\big)\left(4 e_d \tilde{d}_d-e_u \tilde{d}_u\right)+\left(4 d_d-d_u\right)\,,\nn \\
\Theta_p\left(\bar{\theta}, \tilde{d}_q, d_q\right)&=  \chi m_*\left[\left(4 e_u-e_d\right)\left(\bar{\theta}-\frac{m_0^2}{2} \frac{\tilde{d}_s}{m_s}\right)+\frac{m_0^2}{2}\left(\tilde{d}_u-\tilde{d}_d\right)\left(\frac{4 e_u}{m_d}+\frac{e_d}{m_u}\right)+  \frac{m_0^2}{2}\,\left(\frac{4\,e_u}{m_s}\tilde{d}_u-\frac{e_d}{m_s}\tilde{d}_d\right)\right]+ \nn\\
&\hspace{9 cm}+\big(\kappa-\frac{1}{2}\xi\big)\left(4 e_u \tilde{d}_u-e_d \tilde{d}_d\right)+\left(4 d_u-d_d\right)\, 
 .\label{Eq:AnalyticsnpEDMHisano2}
\end{align}\end{widetext}
where $m_* \equiv\big( 1/m_u+1/m_s+1/m_s\big)^{-1}$, $e_u$ and $e_d$ denote the electromagnetic charges of the up and down quarks ($e_u=+2/3\,  e\,,\,e_d=-1/3\,  e$) and the parameters $m_0^2=0.8(2) \, \mathrm{GeV}^2,$ $\kappa=-0.34(10)$, $\chi=-5.7(6) \, \mathrm{GeV}^{-2}$~\cite{Belyaev:1982sa,Ioffe:1983ju} and $ \xi= -0.74(20)$~\cite{Khatsymovsky:1990jb,Kogan:1991en} 
are as defined in Eqs.~(48-50) in Ref.~\cite{Hisano:2012cc}. 
 Substituting th{}ese input parameters in \cref{Eq:AnalyticsnpEDMHisano1,Eq:AnalyticsnpEDMHisano2} we finally obtain: 
\begin{align}
d_n(\bar{\theta}, \tilde{d}_q, d_q)&={0.6(3)\times 10^{-16}\,\bar{\theta}[e\cdot\mathrm{cm}]}\label{Eq:nEDMthetaNumFromHisAnalytics}
\\
&%\hspace{0.2 cm} 
-0.2(1)\,d_u 
+0.8(4)\,d_d
- 0.0028(17)\,d_s \nonumber \\ 
&%\hspace{1cm}
{- 0.32(16)\,e\,\tilde{d}_u
+0.32(16)\,e\,\tilde{d}_d 
- 0.014(7)\,e\,\tilde{d_s}\,,}\nn\\ 
d_p(\bar{\theta}, \tilde{d}_q, d_q)&=-1.0(5) \times 10^{-16}\,\bar{\theta}[e\cdot\mathrm{cm}]\label{Eq:nEDMthetaNumFromHisAnalyticsp}\\
&+ 0.8(4)\,d_u -0.2(1)\,d_d- 0.0028(17)\,d_s \nonumber \\ 
&{-0.26(13)\,e\,\tilde{d}_u  +0.28(14)\,e\,\tilde{d}_d + 0.02(1)\,e\,\tilde{d}_s }\,.\nn,
\end{align}
Note that the coefficients of the qEDMs predicted from these QCD sum rules, with the updated the input values we use, are in good agreement with the lattice QCD result in \cref{Eq:Latice nums qEDM}.  The difference between our numerical results in \cref{Eq:nEDMthetaNumFromHisAnalytics} and 
(\ref{Eq:nEDMthetaNumFromHisAnalyticsp}) and those in \cite{Hisano:2012cc} result solely because we use an updated value for the quark condensate. Note that the apparent discrepancy in the powers of $m_{n}$ in \cref{Eq:AnalyticsnpEDMHisano1} with respect to the expression in Ref.\cite{Hisano:2015rna} results from the fact that the definition of the $c_0$ coefficient in Ref.~\cite{Hisano:2012cc} (and our results) differs by a factor of $m_n^4$ with respect to that in Ref.~\cite{Hisano:2015rna}.

\section {nEDM estimates in the presence of a Peccei-Quinn symmetry} 
\label{App:Bounds with PQ}

The presence of a Peccei-Quinn symmetry (with its corresponding QCD axion) modifies the nEDM estimates in \cref{Eq:nEDMthetadudd}. Firstly, the vev of the QCD axion will cancel the direct $\bth$ dependence in the first term of \cref{Eq:nEDMthetadudd}. Secondly, the presence of explicit CP violation, e.g. via dipole moment couplings,   
 shifts  the minimum of the QCD axion potential away from the CP-conserving minimum, generating an induced $\bth_{\rm ind}$~\cite{Bigi:1991rh,Pospelov:1999ha,Pospelov:1997uv,Pospelov:2000bw}. Indeed, for a chiral CP-odd theory such as that under consideration in \cref{Eq:Initial Lagrangian}, the Vafa-Witten~\cite{Vafa:1983tf} theorem does not apply and therefore cannot  guarantee that the minimum of the axion sits exactly at the CP-conserving minimum. The value of $\bth_{\rm ind}$ is dominated by the cEDM~\cite{Pospelov:1997uv},
\begin{align}
\bth_{\text {ind }}=\frac{m_0^2}{2} \sum_{q=u, d, s} \frac{\tilde{d}_q}{m_q}\,, 
\label{Eq:ThetaInduced}
\end{align}
where our sign convention for $m_0^2$ is that of~\cite{belyaev1982}.
Taking this induced $\bar{\theta}_\text{ind}$ into account results into a modification of the coefficients in \cref{Eq:nEDMthetadudd,Eq:pEDMthetadudd} which now read~\cite{Pospelov:2005pr,Cirigliano:2019vfc} 
\begin{align}
d_n^{\rm PQ} & =-0.204(11)d_u +0.784(28)d_d\nn\\
&\hspace{1 cm}-0.31(15)e\,\tilde{d}_u  + 0.62(31)\,e\,\tilde{d}_d
 \,,
\label{eq:nEDM_diLuzio}\\
d_p^{\rm PQ} & =0.784(28)d_u -0.204(11)d_d\nn \\
&\hspace{1 cm}-1.21(60)\,e\,\tilde{d}_u  -0.15(7)\,e\,\tilde{d}_d
\label{eq:pEDM_PQ}
\end{align}
This result was obtained by substituting, in the estimation of \cref{Eq:nEDMthetadudd}, $\bth$  by the expression for $\bth_{\rm ind}$ in \cref{Eq:ThetaInduced},  i.e. $d_{n,p}^{\rm PQ}(d_{u,d},\tilde d_{u,d})=d_{n,p}(\bth=\bth_{\rm ind}, d_{u,d},\tilde d_{u,d})$. It  differs from the result in Ref.~\cite{Pospelov:2005pr} because we are using a more refined estimation of $d_{n,p}(\bth)$ \cite{Guo:2012vf}. 
Note that in these two equations the dependence on the qEDMs did not suffer any modification with respect to the estimation without PQ symmetry  in Eqs.~(\ref{Eq:nEDMthetadudd}) and (\ref{Eq:pEDMthetadudd}) since the induced $\bth_{\rm ind}$ due to the existence of the qEDM is negligible~\cite{Pospelov:1997uv}, while the impact on the cEDMs is significant as earlier explained. 

The  resulting bounds on ALP-fermion CP-odd couplings  are shown in Table~\ref{table:diLuzioInDerivative} for the case of an ALP with  derivative couplings, and in Table~\ref{table:diLuzioInDerivative-2}  for the case of a general scalar. In the latter case no bounds results on the $\left|\operatorname{Im}(\bld{F}_{q}^{ii})/\Lambda^2\right|$ coefficients, as their contribution is completely reabsorbed away by the Peccei-Quinn symmetry.

\begin{table}[h!]
\centering
\begin{tabular}{ccc}
\multirow{2}{*}[1.1mm]{\text{Combination}}& \text{With PQ}& \multirow{1}{*}{\text{Without PQ}}\\[-1mm]
&  \text{{\small(GeV$^{-2}$)}} & \text{{\small(GeV$^{-2}$)}}    \\[0mm]
\hline\\[-10pt]
 $\left|\operatorname{Im}(\bld{C}_Q^{13}\bld{C}_{u_R}^{13})/f_a^2\right|$
 & $3.7 \times 10^{-12}$
 & $3.7 \times 10^{-12}$
 \\[2mm]
 $\left|\operatorname{Im}(\bld{C}_Q^{13}\bld{C}_{d_R}^{13})/f_a^2\right|$
 & $1.9 \times 10^{-9}$
 & $3.2\times 10^{-9}$
 \\[2mm]
 $\left|\operatorname{Im}(\bld{C}_Q^{12}\bld{C}_{u_R}^{12})/f_a^2\right|$
 & $2.3 \times 10^{-9}$
 & $2.4\times 10^{-9}$
 \\[2mm]
 $\left|\operatorname{Im}(\bld{C}_Q^{23}\bld{C}_{d_R}^{23})/f_a^2\right|$
 & $8.7\times 10^{-9}$
 & $1.2\times 10^{-7}$
 \\[2mm]
 $\left|\operatorname{Im}(\bld{C}_Q^{12}\bld{C}_{d_R}^{12})/f_a^2\right|$
 & $1.2\times 10^{-5}$
 & $1.9\times 10^{-6}$
 \\[2mm]
 \hline \hline
 \end{tabular}
 \vspace{2mm}
 \caption{ {\it ALP case.} Comparison of bounds w/o the presence of a PQ symmetry. All  bounds are in units of GeV$^{-2}$, and  $m_a = 5$ GeV has been assumed for illustration.
 %TO BE REWORKED OUT. WE SHOULD HAVE SOMEWHERE THE DIRECT EDM BOUNDS ASSUMING PQ (and maybe comment lightly on possible differences with di Luzio): }All the bounds are in GeV$^{-2}$ and assume $m_a = 5$\,GeV. Di Luzio's results ASSUME A PQ SYMMETRY, but on the coefficients in the derivative basis. The second column displays di Luzio's reported bounds. \VICTOR{The third column displays my results taking into consideration the the cEDMs ($\tilde{d}_q$'s) only. The fourth column displays my results taking into account also the EDMs ($d_q$'s), as I believe one should do.}
 }
 \label{table:diLuzioInDerivative}
 \end{table}

\begin{table}[h!]
\centering
\begin{align*}
\begin{array}[c]{ccc}
\multirow{2}{*}[1.1mm]{\text{Combination}}& \text{With PQ}& \multirow{1}{*}{\text{Without PQ}}\\[-1mm]
&  \text{{\small(GeV$^{-2}$)}} & \text{{\small(GeV$^{-2}$)}}    \\[0mm]
\hline\\[-10pt]
\left|\operatorname{Im}(\bld{K}_{u}^{13} \bld{K}_{u}^{31})/\Lambda^2\right|
& 9.0 \times 10^{-7}
& 9.2\times 10^{-7}
\\[1mm]
\left|\operatorname{Im}(\bld{K}_{d}^{13} \bld{K}_{d}^{31})/\Lambda^2\right|
& 2.8\times 10^{-7}
& 4.6\times 10^{-8}
\\[1mm]
\left|\operatorname{Im}(\bld{K}_{u}^{12} \bld{K}_{u}^{21})/\Lambda^2\right|
& 3.1 \times 10^{-8}
& 3.1\times 10^{-8}
\\[1mm]
\left|\operatorname{Im}(\bld{K}_{d}^{23} \bld{K}_{d}^{23})/\Lambda^2\right|
& 1.3\times 10^{-6}
& 1.8\times 10^{-5}
\\[1mm]
\left|\operatorname{Im}(\bld{K}_{d}^{12} \bld{K}_{d}^{21})/\Lambda^2\right|
& 8.2 \times 10^{-7}
& 1.4 \times 10^{-7}
\\[1mm]
\left|\operatorname{Im}(\bld{K}_{d}^{22} \bld{K}_{d}^{22})/\Lambda^2\right|
& 3.8\times 10^{-6}
& 5.3 \times 10^{-5}
\\[1mm]
\left|\operatorname{Im}(\bld{K}_{u}^{11} \bld{K}_{u}^{11})/\Lambda^2\right|
& 2.2\times 10^{-6}
& 2.2\times 10^{-6}
\\[1mm]
\left|\operatorname{Im}(\bld{K}_{d}^{11} \bld{K}_{d}^{11})/\Lambda^2\right|
& 8.7\times 10^{-6}
& 1.4 \times 10^{-6}
\\[1mm]
\hline
\hline
\end{array}
\end{align*}
 \vspace{2mm}
 \caption{{\it General scalar.}  Comparison of bounds w/o the presence of a PQ symmetry through the (c)qEDMs. All  bounds are in units of GeV$^{-2}$, and  for $m_\phi = 5$ GeV.
 }
 \label{table:diLuzioInDerivative-2}
 \end{table}

\section{Leading Logs in the quark mass running}
\label{app-running}
Consider  integration of \cref{Eq:deltathetaALPresult1} retaining dependence on the QCD coupling, $\alpha_s$, so that the effect of running masses is retained. We neglect electroweak interactions since their effect is much smaller. 
  To incorporate these effects we solve \cref{Eq:deltathetaALPresult1} making use of the explicit derivative
  \begin{equation}
    \label{mu-deriv}
    \mu \frac{d}{d\mu}= \mu \frac{\partial}{\partial\mu}
    +\beta(g_s)\frac{\partial}{\partial
      g_s}+\sum_qm_q\gamma_m(g_s)\frac{\partial}{\partial 
    m_q}
    \end{equation}
    and the observation that  the couplings $\mathbf{C}_{Q,u_R,d_R}$   have vanishing anomalous dimensions, because  the corresponding operators in the Lagrangian of \cref{Eq:Initial Lagrangian} are partially conserved currents in QCD. Denoting by $\bar g$ and $\bar m$ the running coupling and quark mass in QCD, that is the solutions to
      \[
        \mu\frac{d\bar g}{d\mu}=\beta(\bar g)\quad\text{and}\quad
\mu\frac{d\overline m}{d\mu}=\gamma_m(\bar g)\overline m\,,
\]
integration of \cref{Eq:deltathetaALPresult1} including the full $\mu$-derivative of \cref{mu-deriv} is equivalent to  using only the $\mu\partial\bth/\partial \mu$ on the left hand side and substituting  running couplings for the couplings on the right hand side.

 Leading-log resummation is achieved using one-loop running
 couplings. Using
 \[
   \beta(g)=-b_0\frac{g^3}{16\pi^2}\quad\text{and}\quad
   \gamma_m(g)=a_m\frac{g^2}{16\pi^2}
 \]
 with $b_0  = 11 - 2 n_f/3$ and $a_m = -8$ , the running couplings $\overline\alpha_s(\mu)$
 are 
   \begin{equation}
     \label{eq:runalpha}
\frac1{\overline \alpha_s(\mu)}-\frac1{\overline \alpha_s(\mu')}=\frac{b_0}{2\pi}\ln(\mu/\mu')
   \end{equation}
   and 
   \begin{equation}
     \label{eq:runmass}
     \overline m(\mu)=\overline m(\mu')\left(\frac{\bar \alpha_s(\mu)}{\bar \alpha_s(\mu')}\right)^{-a_m/2b_0}\,.
   \end{equation}
Notice that the running of the  factor of $\mathbf{M}^{-1}$ on the right hand side on  \cref{Eq:deltathetaALPresult1} cancels with the
explicit overall mass factor in $\mathcal{L}$ of \cref{Eq:deltathetaALPresult2}. What remains is an integral over $m_a^2+m_q^2$. The first of these is $\mu$-independent so that integration simply gives a logarithm of the ratio of scales, as in \cref{final-result-theta-BB}. For the second term one needs 
\begin{widetext}

\begin{align*}
\int_{\mu'}^\mu\frac{d\mu}{\mu}\overline
  m^2=\int_{g'}^g\frac{d\bar g}{\beta(\bar g)}\overline m(\bar g)^2
  =-\frac{16\pi^2}{b_0}\overline m^2(\mu') \int_{g'}^g\frac{d\bar
    g}{\bar g^3}\left(\frac{g'}{\bar g}\right)^{2a_m/b_0}
 = \frac{8\pi^2}{b_0(1+a_m/b_0)}\left(\frac{\overline m^2(\mu)}{\bar g^2(\mu)}-\frac{\overline m^2(\mu')}{\bar g^2(\mu')}\right)\,.
\end{align*}
In the limit of vanishing coupling, $g\to0$ this gives $m^2\ln(\mu/\mu_0)$, reproducing the  explicit log in
  \cref{final-result-theta-BB}. Accordingly one should use
\begin{equation}
\hat
m^2_{q_k}\log(\mu_0^2/\mu_1^2)=\frac{4\pi}{b_0(1+a_m/b_0)}\left(\frac{\overline m^2(\mu_0)}{\overline \alpha_s (\mu_0)}-\frac{\overline
    m^2(\mu_1)}{\overline \alpha_s(\mu_1)}\right)\,,
\label{hat-mass-given}
\end{equation}
with $\mu_0=f_a$ and $\mu_1 = \texttt{max}(m_a^2,m_{d_k}^2)$ in \cref{final-result-theta-BB}.

A couple of remarks are in order. First, when computing the ratios  $m_{u_k}/m_{u_i}=\overline m_{u_k}(\mu)/\overline
  m_{u_i}(\mu)$ and $m_{d_k}/m_{d_i}=\overline m_{d_k}(\mu)/\overline  m_{d_i}(\mu)$ at a common renormalization scale $\mu$, the running of  the mass and  coupling constant may change as they cross through thresholds. For example, for the $m_{u_k}/m_{u_i}=m_t/m_u$ case one writes the ratio in terms of $\overline m_t(m_t)$ and $\overline  m_u(\mu_{IR})$ with $\mu_{IR}=\text{2\;GeV}$, the values at the scales where they are  often reported, as
  \[
    \frac{m_t}{m_u}
    =\left(\frac{\overline\alpha^{(5)}_s(m_t)}{\overline\alpha^{(5)}_s(m_b)}\right)^{\frac{12}{23}}\left(\frac{\overline\alpha^{(4)}_s(m_b)}{\overline\alpha^{(4)}_s(\mu_{IR})}\right)^{\frac{12}{25}}\frac{\overline m_t(m_t)}{\overline m_u(\mu_{IR})}
    \]   
    where $\overline\alpha_s^{(n_f)}(\mu)$ is the running coupling of \cref{eq:runalpha} computed with $n_f$ active  quark flavours.  And second, similarly, the effective mass in \cref{hat-mass-given} accounts for threshold effects additively. For example, if $m_a<m_b$ one should use
   \begin{multline}
     \widehat
m^2_{b}\log(\mu_0^2/m_b^2)=4\pi\overline m_b^2(m_b)
\left[
  \left(\frac{\overline\alpha^{(5)}_s(m_t)}{\overline\alpha^{(5)}_s(m_b)}\right)^{\frac{24}{23}}
  \left(\frac{1}{\overline \alpha^{(6)}_s(m_t)}-
    \frac{1}{\overline
      \alpha_s^{(6)} (\mu_0)}\left(\frac{\overline\alpha^{(6)}_s(\mu_0)}{\overline\alpha^{(6)}_s(m_t)}\right)^{\frac{24}{21}} \right)\right.\\
\left.  +
  3\left(\frac{1}{\overline
      \alpha^{(5)}_s(m_b)}-\frac{1}{\overline
      \alpha_s^{(5)} (m_t)}\left(\frac{\alpha^{(5)}_s(m_t)}{\alpha^{(5)}_s(m_b)}\right)^{\frac{24}{23}}\right)
\right]\,.
\label{hat-mass-given-mb}
\end{multline}
\end{widetext}

To get a sense of the magnitude of the leading log (LL) resummation we take $m_a<m_t$ and  $f_a=10^8\;\text{GeV}$ which would be the minimum  required from our conservative bounds in \cref{Tab: bounds for different couplings} for $\left|\operatorname{Im}[\bld{C}_{Q}^{13}\bld{C}_{u_R}^{*13}]\right|\sim10^{-2}$. Then we use, naively, $\widehat m_q=\overline m_q(m_q)$ and then compare both
sides of \cref{hat-mass-given} and of \cref{hat-mass-given-mb}. Using $\overline \alpha_s^{(5)}(M_z)=0.12$, $m_t=175\;\text{GeV}$,
  $m_b=4.2\;\text{GeV}$ we obtain that the naive log overestimates the LL by  factors of 1.8 and 2.8 in  \cref{hat-mass-given} and
  \cref{hat-mass-given-mb}, respectively. It is a curious coincidence  that the running of the up-quark from 1.0\;GeV to $m_t$ required to
  compute the factor $m_t(m_t)/m_u(m_t)$ in \cref{final-result-theta-BB} is 1.8, exactly compensating for the LL
  resummation:
  \[
    \frac{\overline m_t(m_t)}{\overline m_u(1\;\text{GeV})}\overline m_t^2(m_t)
  \approx  \frac{\overline m_t(m_t)}{\overline m_u(m_t)}
    \widehat m_t^2
  \]
  This coincidence is not generic; the running of the $d$ or $s$ quark masses from 1.0\;GeV to $m_b$ gives an enhancement of 1.3 to
  the ratio $m_b/m_{d,s}$, so the overall LL resummation effect is a a factor of 2.1 suppression.

  We hasten to remind the reader that the bounds on couplings in \cref{sec:impact} use, conservatively, $\widehat m_q^2\log(f_a^2/\mu_1^2)\to m_q^2$, with $m_q= \overline m_q(m_q)$  for $q=t,b,c$ and $\overline m_q(1\;\text{GeV})$ for $q=u,d,s$. For the parameters chosen in the examples above the neglected logarithmic factor is $\sim20$.  
It goes without saying that including this rather large logarithmic factor results in stronger bounds than the ones we obtained.

\bibliographystyle{utphysUCSD}
\bibliography{Bibliography}

\end{document}